\documentclass[print]{elsarticle}
\usepackage{lineno,hyperref}
\usepackage{graphicx}
\usepackage{subfig}
\usepackage[linesnumbered,ruled,lined]{algorithm2e}
\usepackage{float}
\usepackage{tikz}
\usepackage{epstopdf}
\usepackage{amssymb}
\usepackage{amsmath}
\journal{J. Comput. Phys.}
\newcommand{\tr}{\mathop{\mathrm{tr}}}

\newcommand\crule[3][black]{\textcolor{#1}{\rule{#2}{#3}}}
\newcommand{\etal}{\text{et al.}}
 \pdfoutput=1
\begin{document}
	\begin{frontmatter}
		\title{A multi-resolution SPH framework: Application to multi-phase fluid-structure interactions}
		\author[myfirstaddress]{Chi Zhang }
		\ead{c.zhang@tum.de}
		\author[myfirstaddress,mysecondaryaddress]{Yujie Zhu}
		\ead{yujie.zhu@tum.de}
		\author[myfirstaddress]{Xiangyu Hu \corref{mycorrespondingauthor}}
		\ead{xiangyu.hu@tum.de}
		\address[myfirstaddress]{TUM School of Engineering and Design, Technical University of Munich, 85748 Garching, Germany}
		\address[mysecondaryaddress]{Xi'an research institute of Hi-Tech, 
			70025 Xi'an, China}
		\cortext[mycorrespondingauthor]{Corresponding author.}
	\begin{abstract}
		In the previous work, 
		Zhang \etal~(2021)~\cite{zhang2021multi} proposed a multi-resolution smoothed particle hydrodynamics (SPH) method 
		for fluid-structure interactions (FSI) with achieving an optimized computational efficiency meanwhile maintaining good numerical robustness and accuracy. 
		In the present paper, 
		this multi-resolution SPH framework where different spatial-temporal discretizations are applied for different sub-systems
		is extended to multi-phase flows involving large density ratio and interacting with rigid or flexible structure. 
		To this end,  
		a simple and efficient multi-phase model is introduced 
		by exploiting different density reinitialization strategies other than applying different formulations to implement mass conservation 
		to the light and heavy phases, respectively, 
		to realize the target of using same artificial speed of sound for the both. 
		To eliminate the unnatural voids and unrealistic phase separation meanwhile decrease the numerical dissipation, 
		the transport velocity formulation are rewritten by applying temporal local flow state dependent background pressure. 
		To handle the FSI coupling in both single- and multi-resolution scenarios in the triple point, 
		the one-sided Riemann-based solid boundary condition is adopted. 
		A set of examples involving multi-phase flows with high density ratio and complex interface 
		and multi-phase FSI are studied to demonstrate the efficiency, accuracy and robustness of the present method. 
		The validations presented herein and those reported in the original paper of Zhang \etal~(2021)~\cite{zhang2021multi} 
		where single-phase FSI is studied 
		put the present multi-resolution SPH framework in good stead in terms of computational efficiency for multi-physics applications. 
	\end{abstract}
	\end{frontmatter}
%
%
\clearpage
\section{Introduction}\label{sec:introduction}
Multi-phase flow is ubiquitous in natural phenomena, 
e.g. rain drops, soap bubbles and water beading on a leaf. 
Extensive studies have been conducted in the past decades to unravel the underlying physics, 
and to address the associated large variety of engineering applications, 
e.g. 
thermal spray coating, ink-jet printing and multi-component flow in the chemical reactors. 
Many of these studies have focused on multi-phase flow involving large density ratio and interacting with rigid or flexible structures. 
However, 
numerical study is highly challenging 
due to the intrinsic density discontinuity and complex physical interactions across the dynamically evolving material interface. 
As a promising alternative of the Eulerian mesh-based methods \cite{hirt1981volume, sussman1994level, hu2006conservative}, 
the fully Lagrangian and mesh-free smoothed particle hydrodynamics (SPH), 
which was originally developed by Lucy \cite{lucy1977numerical} and Gingold and Monaghan \cite{gingold1977smoothed} for astrophysical applications, 
has shown peculiar advantages in handling multi-phase flows and fluid-structure interactions (FSI) 
\cite{monaghan1995sph, colagrossi2003numerical, hu2006multi, zhang2015sph, sun2019study}. 
Thanks to its very fully Lagrangian nature, 
the SPH method shows instinctive advantage in handling material interface without 
additional interface tracking or capturing algorithms \cite{tryggvason2001front, pan2018high} 
which suffers from serious numerical errors and instabilities and induces extra computational efforts, 
especially when flexible structure displaced in multi-phase flow is involved \cite{onate1996finite}. 

Concerning the modeling of incompressible multi-phase flow and its interaction with rigid or flexible structure, 
the weakly-compressible (WC) assumption \cite{monaghan1992smoothed}, 
where the fluid pressure is calculated from the density through an equation of state (EoS) with proper choice of artificial speed of sound, 
is most widely applied in literature \cite{colagrossi2003numerical, hu2006multi}. 
Notwithstanding its versatile application and promising achievement, 
the WC-assumption induces spurious pressure oscillations, 
leading to numerical instability which exhibits unnatural voids and unrealistic phase separation 
in the simulation of multi-phase flow with high density ratio and violent interface change\cite{colagrossi2003numerical,chen2015sph, rezavand2020weakly}. 
To remedy this issue, 
tremendous efforts have been conducted in the literature. 
Colagrossi and Landrini \cite{colagrossi2003numerical} proposed a multi-phase SPH method by adopting 
Mean Least Square (MLS) interpolation-based density reinitialization, 
using large surface tension for light phase and implementing XSPH correction \cite{monaghan1995sph} to realize isotropic particle distribution. 
This method  can achieve stable modeling of multi-phase flows involving high density ratio and interacting with rigid structure,  
while greatly changes the natural physical property of the fluid \cite{chen2015sph} 
and introduces excessive computational costs by assigning the light phase 
$10$ times larger artificial speed of sound than the one of the heavy phase \cite{rezavand2020weakly}. 
Following Ref. \cite{colagrossi2003numerical}, 
efforts were devoted to avoiding unnatural void by applying particle shifting algorithm \cite{mokos2017multi} 
and to produce smooth pressure field by introducing diffusive term in the continuity equation \cite{antuono2012numerical, hammani2020detailed}.
However, 
the issue of being excessive computationally expensive still hinders its further applications in large scale simulation. 
Hu and Adams \cite{hu2006multi} proposed a multi-phase SPH method for both macroscopic and mesoscopic flows by 
deriving the particle-averaged spatial derivative approximation from 
a particle smoothing function in which the neighboring particles only contribute to the specific volume, while maintaining mass conservation. 
This method can naturally handle the density discontinuity across phase interface and achieve efficient simulation by assigning identical artificial speed of sound for both phases, while its applications to complex and violent interface is not fully addressed. 
Similar with Ref. \cite{hu2006multi}, 
Grenier \etal \cite{grenier2009hamiltonian} proposed a Hamiltonian interface SPH formulation 
and introduced small repulsive force at the interface to capture sharp interface evolution.
This method still suffers from being low computational efficiency due to the large artificial speed of sound for light phase 
and requires extra sweeping and computational efforts for determining volume distribution of particle. 
Similar with Ref. \cite{grenier2009hamiltonian}, 
Monaghan and Rafiee \cite{monaghan2013simple} 
developed a simple multi-phase SPH formulation by rewriting the repulsive force as 
the artificial stress formulation \cite{monaghan2000sphtensile}. 
This method slightly improves the computational efficiency by reducing the artificial speed of sound for the light phase 
to the $4$ times value of that for the heavy phase. 
Different with Ref. \cite{colagrossi2003numerical}, 
Chen \etal \cite{chen2015sph} developed a multi-phase SPH method with the assumption of pressure and space continuities at the interface, 
and introduced corrected density reinitialization and positive background pressure to obtain smooth and positive pressure field. 
This method can achieve optimized computational efficiency as same speed of sounds is realized, 
while exhibits excessive numerical dissipation \cite{chen2015sph, rezavand2020weakly}.
Zheng and Chen \cite{zheng2019multiphase} proposed a further modification with a first-order density reinitialization and a local
implementation of the artificial viscosity. 
This modification demonstrates enhanced accuracy, such as shorter delays of the
dynamics, due to less numerical dissipation. 
However, 
particles still undergo spurious fragmentation, 
particularly in the vicinity of phase interface in their violent water-air flow cases. 
More recently, 
Rezavand \etal \cite{rezavand2020weakly} proposed a new mutli-phase SPH method by 
exploiting the two-phase Riemann solver to handle the pairwise particle interaction 
and applying the transport-velocity formulation to the light phase. 
This  mutli-phase method assumes that the light phase experiences the heavy phase
like moving boundary, while the heavy phase undergoes a free-surface-like flow.
This method can achieve optimized computational efficiency by using the same value for the artificial speed of sound in both phases, 
meanwhile effectively eliminates the unnatural voids and unrealistic phase separation. 

In addition to pursuing efficient multi-phase SPH simulation by applying the same artificial speed of sound in both phases, 
the computational efficiency can be further optimized by adopting multi-resolution scheme. 
Since its inception, 
the SPH method has experienced tremendous progresses in developing 
different accurate, stable and consistent multi-resolution schemes which can be generally classified into four classes:
adaptive particle refinement (APR) with or without particle splitting/merging 
\cite{springel2005cosmological,lastiwka2005adaptive, vacondio2016variable, khorasanizade2016dynamic, hu2019consistent, yang2021a}, 
non-spherical particle scheme \cite{liu2006adaptive, owen1998adaptive}, 
domain-decomposition based scheme \cite{bian2015multi, shibata2017overlapping, khayyer2019multi, zhang2021multi} 
or the hybrid scheme \cite{barcarolo2014adaptive, tanaka2018multi, omidvar2012wave}. 
However, 
implementing multi-resolution scheme to particle-based 
incompressible multi-phase flow and FIS simulations has only emerged in more recent years. 
Yang \etal \cite{yang2019adaptive, yang2021b} proposed an adaptive spatial resolution (ASR) method \cite{yang2021a} 
for multi-phase flows by  varying the spatial resolution with respect to the distance to the interface and 
exploiting particle shifting technique with the consideration of variable smoothing length. 
Sun \etal \cite{sun2019study} focused on the multi-phase hydroelastic FSI problems by adopting 
the multi-resolution scheme developed by Barcarolo et al. \cite{barcarolo2014adaptive}. 
Zhang \etal \cite{zhang2022} studied the multi-phase FSI problem 
by exploiting multi-resolution discretization strategy proposed by Zhang \etal \cite{zhang2021multi} 
in fluid-structure interface 
while resolving the fluid interface by using the multi-phase model developed by Rezavand \etal \cite{rezavand2020weakly} 
in single-resolution scenario. 
However,  
applying the multi-resolution framework of Ref. \cite{zhang2021multi} to multi-phase flows with large density ratio, 
in particular involving interaction with flexible structures, 
is still not addressed. 

In this work, 
the multi-resolution framework developed by Zhang \etal \cite{zhang2021multi} 
where different spatial-temporal discretizations are applied for different sub-systems
is extended to multi-phase flows involving large density ratio and interacting with rigid or flexible structures. 
In cooperate with this multi-resolution framework, 
a simple and efficient multi-phase model is proposed by introducing different density reinitialization strategies 
other than applying different formulations to implement mass conservation as in Ref. \cite{rezavand2020weakly}  
for the light and heavy phases to realize the target of using same artificial speed of sound for the both.
To decrease the numerical dissipation meanwhile preserve the feature of eliminating the unnatural voids and unrealistic phase separation, 
the transport-velocity formulation is rewritten by introducing a background pressure based on the temporal reference of current flow state.  
To handle the FSI coupling with both single- and multi-resolution discretizations present,  
the one-sided Riemann-based solid boundary condition \cite{zhang2017weakly, zhang2022} is adopted. 
A set of examples, 
e.g. multi-phase hydrostatic test, 
multi-phase flows with high density ratio and complex interface and multi-phase flow interacting with flexible structures,
are studied to demonstrate the efficiency, accuracy and robustness of the present method. 
The numerical algorithms are implemented in the open-source SPHinXsys library \cite{zhang2020sphinxsys, zhang2021sphinxsys}, 
and all the computational codes and data-sets accompanying this work are available at \url{https://www.sphinxsys.org}. 
This paper is organized as follows. 
Section \ref{sec:governing} briefly summarizes the governing equations for both fluid and solid dynamics 
and Section \ref{sec:methodology} presents the detailed methodology.
Then, 
the numerical validations are presented and discussed in Section \ref{sec:examples}, 
and concluding remarks are finally given in Section \ref{sec:conclusion}. 
%
%
\clearpage
\section{Governing equations}\label{sec:governing}
In this paper, 
we consider 
the viscous and immiscible two-phase flow with large density ratio and its interaction 
with rigid or flexible structure which may experience large deformation. 
For the two-phase flow with large density ratio, 
i.e., $ \rho_l/\rho_g \gg 1$ with $\rho_l$ and $\rho_g$ denoting the densities of light and heavy phase respectively, 
the mass and momentum conservation equations read
\begin{equation} 
\begin{cases}\label{eq:fluid-governing}
\frac{\text d \rho}{\text d t}  =  - \rho \nabla \cdot \mathbf v \\
\frac{\text d \mathbf v}{\text d t}  =   \frac{1}{\rho} \left[ - \nabla p +  \eta \nabla^2 \mathbf v \right] 
+ \mathbf g + \mathbf f^{s:p} + \mathbf f^{s:\nu} ,
\end{cases}
\end{equation}
where $\frac{\text d}{\text d t}=\frac{\partial}{\partial t} + \mathbf v \cdot \nabla$ is the material derivative, 
$\rho$ the density, 
$\mathbf v$ the velocity, 
$p$ the pressure, 
$\eta$ the dynamic viscosity, 
$\mathbf g$ the gravity, 
and $\mathbf f^{s:p}$ and $\mathbf f^{s:\nu}$ respectively represent the pressure and viscous force acting on the fluid 
due to the presence of rigid or flexible solid. 
With the weakly-compressible assumption \cite{morris1997modeling} in mind,  
a linear equation of state (EoS)
\begin{equation}\label{eq:EoS}
p = {c^f}^2 (\rho-\rho^0),
\end{equation} 
for both the heavy and light phases is introduced to close the system of Eq. \eqref{eq:fluid-governing}. 
Here, 
$\rho^0$ is the initial reference density 
and $c^f = 10U_{max}$ the artificial speed of sound with $U_{max}$ denoting the 
maximum anticipated flow speed. 
Note that the compressible effect of the light phase is neglected following 
Refs. \cite{flekkoy2000foundations, hu2006multi,grenier2009hamiltonian, zhang2015sph, rezavand2020weakly}. 
Also, 
identical value for the artificial speed of sound is applied for both phases to optimize the computational efficiency 
as Ref. \cite{rezavand2020weakly}. 

For the flexible solid, 
we consider an elastic and weakly-compressible material. 
With the definition of the initial position $\mathbf{r}^0$ and the current position $\mathbf{r}$ 
of a material point in the initial reference and the deformed current configuration, respectively, 
the deformation tensor $\mathbb F$ can be defined by
\begin{equation}\label{eq:deformation-tensor}
\mathbb F  = \nabla^0 \mathbf u  + \mathbb I,
\end{equation}
where $\nabla^0 \equiv \frac{\partial}{\partial \mathbf r^0}$ 
stands for the gradient operator with respect to the initial reference configuration, 
$\mathbf u = \mathbf r - \mathbf r^0 $ the displacement and 
$\mathbb I$ the identity matrix.  
Here the superscript $\left( {\bullet} \right)^0$ is introduced to denote the quantities in the initial reference configuration.
Then,  
the mass and momentum conservation equations are defined following total Lagrangian framework as 
\begin{equation} 
\begin{cases}\label{eq:solid-governing}
\rho  =  \rho^0 \frac{1}{J} \\
\frac{d \mathbf v}{d t}  = \frac{1}{\rho^0 } \nabla^0 \cdot \mathbb P^T + \mathbf f^{f:p} + \mathbf f^{f:\nu} ,
\end{cases}
\end{equation}
where $J = \det(\mathbb{F})$ is the Jacobian determinant of deformation tensor $\mathbb{F}$, 
$\mathbb P  = \mathbb{F} \mathbb{S}$ the first Piola-Kirchhoff stress tensor with $T$ denoting transpose operation 
and $\mathbb{S}$ the second Piola-Kirchhoff stress tensor, 
and $\mathbf f^{f:p}$ and $\mathbf f^{f:\nu}$ 
represent the fluid pressure and viscous force acting on the solid respectively. 
To close the system of Eq. \eqref{eq:solid-governing}, 
the constitutive equation for linear elastic and isotropic material reads
\begin{eqnarray}\label{isotropic-linear-elasticity}
	\mathbb{S} & = & K \tr\left(\mathbb{E}\right)  \mathbb{I} + 2 G \left(\mathbb{E} - \frac{1}{3}\tr\left(\mathbb{E}\right)  \mathbb{I} \right) \nonumber \\
	& = & \lambda \tr\left(\mathbb{E}\right) \mathbb{I} + 2 \mu \mathbb{E} ,
\end{eqnarray}
where $\lambda$ and $\mu$ are the Lamé parameters \cite{sokolnikoff1956mathematical}, 
$K = \lambda + (2\mu/3)$ the bulk modulus and $G = \mu$ the shear modulus. 
Note that we consider weakly compressible material whose
artificial speed of sound is defined by $c^{s} = \sqrt{K/\rho}$. 
Also, we introduce $c^{s}$ to denote the sound speed of flexible structure  
to distinguish from the one $c^{f}$ for fluid, viz., both light and heavy phases. 
%
%
\clearpage
\section{Methodology}\label{sec:methodology}
In this section, 
we first present the extension of the multi-resolution framework \cite{zhang2021multi} for multi-phase flow 
with light and heavy phases discrertized by different spatial resolutions using the Riemann-based SPH method. 
Then, 
the multi-phase density reinitialization strategy and the modified transport-velocity formulation are provided. 
Last but not least, 
the total Lagrangian formulation for discretizing solid dynamics 
and the fluid-solid interface treatment are detailed with the time integration as follows.   
\subsection{Multi-phase Riemann-based SPH}\label{sec:fluid-discretization} 
Following the multi-resolution framework where different smoothing lengths are applied to diescretize different sub-systems 
developed in Ref. \cite{zhang2021multi}, 
we introduce $h^l$ and $h^h$ to denote the smoothing length for light and heavy phase discretizations, 
respectively. 
Similar with Ref. \cite{zhang2021multi}, 
we assume that $h^l > h^h$, implying the dynamics of heavy phase is resolved at a higher spatial resolution, 
while the computational efficiency is enhanced when a lower resolution for the light phase is sufficient.
Following the Refs. \cite{zhang2017weakly, rezavand2020weakly}, 
the mass and  momentum conservation equations of Eq. \eqref{eq:fluid-governing} 
can be discretized with the Riemann-based SPH method as 
\begin{equation} 
\footnotesize
\begin{cases}\label{eq:rieammsph-mr}
\frac{\text d\rho_i}{\text d t} = 2\rho_i \sum_j V_j (\mathbf v_i - \mathbf v^\ast) \cdot\nabla_{i} W_{ij}^{h^\chi}\\
\frac{\text d \mathbf v_i}{\text d t} = - \frac{2}{m_i} \sum_j V_i V_j P^\ast \nabla_i W_{ij}^{h^\chi} +  \frac{2}{m_i}\sum_j V_i V_j \frac{2\eta_i \eta_j}{\eta_i + \eta_j} \frac{\mathbf v_{ij}}{r_{ij} }
\frac{\partial W_{ij}^{h^\chi}}{\partial r_{ij}} + \mathbf g + \mathbf f_i^{s:p}\left( h^\chi\right)  + \mathbf f_i^{s:\nu}\left(h^\chi \right) ,
\end{cases} 
\end{equation} 
where $m$ is the particle mass, 
$V$ particle volume, 
$\mathbf v_{ij} = \mathbf v_i - \mathbf v_j$ the relative velocity, 
$\mathbf r$ the particle position, 
$r_{ij} = \left| \mathbf r_i - \mathbf r_j\right|$ the particle distance, 
$\nabla_i W_{ij}^{h^\chi} = 	\frac{\mathbf r_i - \mathbf r_j}{r_{ij}} \frac{\partial W(r_{ij}, h^\chi)}{\partial {r_{ij}}}$ 
the gradient of the kernel function with respect to particle $i$, 
and $\mathbf v^\ast$ and $P^\ast$ are the solutions of an
inter-particle Riemann problem along the unit vector $-\mathbf{e}_{ij} = \frac{\mathbf r_j - \mathbf r_i}{r_{ij}}$.
Note that $h^\chi = h^l$ or $h^\chi = h^h$ is applied for particle interactions inside a single phase, 
i.e., light or heavy phase, 
while $h^\chi = h^l = \max (h^l, h^h)$ for particle interactions at the interface. 
In this case, 
$h^\chi = h^l $ is applied for both light and heavy phase particles located at the interface, 
indicating that the light phase particle $i$ can be searched and tagged 
as a neighboring particle of a heavy phase particle $j$ which is located in its neighborhood.

Following Ref. \cite{zhang2017weakly},  
the initial left and right states of the inter-particle Riemann problem are reconstructed as 
\begin{equation}\label{eq:lrstates}
	\begin{cases}
		(\rho_L, U_L, p_L) = (\rho_i, -\mathbf v_i \cdot \mathbf e_{ij}, p_i)\\
		(\rho_R, U_R, p_R) = (\rho_j, -\mathbf v_j \cdot \mathbf e_{ij}, p_j), 
	\end{cases}
\end{equation}
where subscripts $L$ and $R$ denote left and right states, respectively. 
With applying the dissipation limiter proposed by Zhang et al . \cite{zhang2017weakly}  to the linearized Riemann solver, 
the Riemann solution can be obtained by 
\begin{equation}\label{eq:lrsolution}
	\begin{cases}
		U^\ast = \frac{\rho_L U_L + \rho_R U_R}{\rho_L+\rho_R} +\frac{p_L - p_R}{c(\rho_L+\rho_R)} \\
		p^\ast = \frac{\rho_L p_R + \rho_R p_L}{\rho_L+\rho_R} + \frac{\rho_L\rho_R\beta(U_L - U_R)}{\rho_L+\rho_R},
	\end{cases}
\end{equation}
where the limiter $\beta= \mathrm{min}\left[ 3 \mathrm{max} (U_L - U_R,0), c^f \right] $ 
implying that there is no numerical dissipation imposed when the fluid is under the action of an expansion wave 
and the numerical dissipation is modulated when the fluid is under the action of a compression wave \cite{zhang2017weakly}.
With $U^\ast $ in hand, 
the $\mathbf v^\ast$ can be obtained by 
\begin{equation}
	\mathbf v^\ast = \left( U^\ast - \frac{\rho_L U_L + \rho_R U_R}{\rho_L+\rho_R} \right)  \mathbf e_{ij} + \frac{\rho_i \mathbf{v}_i + \rho_j \mathbf{v}_j}{\rho_i+\rho_j} .
\end{equation}
%
\subsection{Density reinitialization for multi-phase flow}\label{sec:densityreinit}
In Ref. \cite{rezavand2020weakly}, 
a multi-phase model, 
where the light phase experiences the heavy phase like moving wall boundary while the heavy phase undergoes a free-surface-like flow, 
is proposed to eliminate non-physical voids or fragmentation meanwhile 
optimize the computational efficiency by applying the same value for the artificial speed of sound in both phases \cite{colagrossi2003numerical}. 
In particular, 
the density summation is applied in the light phase, 
while the continuity equation is dicretized for the heavy phase, 
to implement mass conservation. 
Inspired by Ref. \cite{rezavand2020weakly}, 
we introduce different density reinitialization strategies, 
while apply identical discretization strategy, 
to the light and heavy phases to realize the multi-phase model of Ref. \cite{rezavand2020weakly}. 
More specifically,  
the continuity equation is discretized by using Eq. \eqref{eq:rieammsph-mr} for both phases, 
while the density field for heavy phase is reinitialized every advection time step \cite{zhang2020dual} with  
the formulation \cite{zhang2022}
\begin{equation} \label{eq:rhosumsurface}
	\rho_i = \rho^0 \frac{ \sum W\left( r_{ij}, {h^h}\right)  }{\sum W^0\left( r_{ij}, {h^h}\right) } 
	       +  \max\left[ 0, ~\rho^* - \rho^0 \frac{ \sum W\left( r_{ij}, {h^h}\right) }{\sum W^0\left( r_{ij}, {h^h}\right)}\right] \frac{\rho^0}{\rho^\ast},
\end{equation}
and for the light phase Eq.  \eqref{eq:rhosumsurface} is rewritten as 
\begin{equation} \label{eq:rhosum} 
	\rho_i =  \rho^0 \frac{ \sum W(r_{ij}, {h^\chi}) }{\sum W^0\left( r_{ij}, {h^\chi}\right)  } . 
\end{equation}
implying that the light phase experiences internal flow. 
Here,  $\rho^\ast$ denotes the density before reinitialization.
Note that the density reinitialization of the heavy phase ignore the presence of the light phase, 
whereas the light phase takes the heavy phase into account with $h^\chi = h^l$ for particle summation in the phase interface. 
\subsection{Transport-velocity formulation}\label{sec:transportvelocity}
Due to the tensile instability \cite{monaghan2000sphtensile,zhang2017generalized}, 
the SPH results particle clumping and unnatural 
void regions in the simulation of fluid dynamics 
when negative pressure presents \cite{adami2013transport}. 
In the simulation of multi-phase flow, in particular large density and complex interface are involved, 
such issue becomes more severe and exhibits unnatural void region and phase separation \cite{mokos2017multi}.  
To address this issue, 
particle shifting or transport-velocity formulations is applied 
in literature \cite{colagrossi2003numerical, sun2019study, rezavand2020weakly, zhang2022}. 
Similar with Ref. \cite{rezavand2020weakly}, 
the transport-velocity formulation \cite{adami2013transport, zhang2017generalized, zhu2021consistency} is only applied on the particles of light phase,
implying that the particle advection velocity $\tilde{\mathbf{v}}$, 
which is used to advect the particles, 
is obtained by 
\begin{equation} \label{eq:transprort}
	\tilde{\mathbf{v}}_i(t + \delta t) = \mathbf{v}_i(t)  + \delta t \left( \frac{\text{d}\mathbf{v}_i}{\text{d}t} 
	- \frac{2}{m_i}\sum_{j} V_i V_j \widehat{p}_0 \nabla_{i} W_{ij}^{h^\chi}\right) , 
\end{equation}
where $\widehat{p}_0$ is the background pressure \cite{adami2013transport, zhang2017generalized}. 
In Ref. \cite{rezavand2020weakly}, 
a spatiotemporal constant background pressure of $\widehat{p}_0  = 4 \rho^0 {c^f}^2$ 
is applied, 
introducing extra artificial damping due to the fact that the transport-velocity formulation 
induces particle freezing. 
In this work, 
we modify the background pressure as $\widehat{p}_0  = 10 \rho_i U^2_{max}$ with $U_{max}$ denoting 
the current maximum particle speed to decrease the artificial damping. 
Inspired by Ref. \cite{zhang2017generalized} where a spatiotemporal localized background pressure is applied, 
the present modification is characterized by being spatial constant while temporal localized. 
The parameter of the background pressure is set according to numerical experiment with the target of maximally reducing 
the artificial damping meanwhile effectively eliminating particle clumping
and applied through all the tests in this paper. 
Note that the present formulation can reduce the particle freeze induced artificial viscosity.  
It is worth noting that the light phase takes the heavy phase into account for applying the transport velocity 
with $h^\chi = h^l$ for particle at the interface. 
\subsection{Total Lagrangian formulation}\label{sec:solid-discretization} 
Following Refs. \cite{zhang2021multi, zhang2021integrative}, 
the discretization of the mass and momentum conservation equations of Eq. \eqref{eq:solid-governing} 
can be written in the total Lagrangian formulation as 
\begin{equation}
\begin{cases}
\rho_a = \rho_a^0 \frac{1}{J} \\
\frac{\text d \mathbf v_a}{\text d t_t} = \frac{1}{m_a} \sum_b V_a V_b \tilde{\mathbb P}_{ab} \nabla_a^0 W_{ab}^{h^s} +\mathbf g 
+ \mathbf f_a^{f:p}\left(h^\chi \right)  +  \mathbf f_a^{f:\nu}\left(h^\chi \right),
\end{cases}
\end{equation}
where inter-particle averaged first Piola-Kirchhoff stress $\tilde{\mathbb{P}}$ reads
\begin{equation}
\tilde{\mathbb P}_{ab} = \frac{1}{2} \left( \mathbb P_a \mathbb B_a^0 + \mathbb P_b \mathbb B_b^0 \right),  
\end{equation}
with adopting  
the correction matrix \cite{vignjevic2006sph, randles1996smoothed, zhang2021multi} 
\begin{equation} \label{correctmatrix}
\mathbb B^0_a = \left( \sum_b V_b^0 \left( \mathbf r_b^0 - \mathbf r_a^0 \right) \otimes \nabla_a^0 W_{ab}^{h^s} \right) ^{-1}. 
\end{equation}
to reproduce rigid-body rotation. 
Here, 
$h^s$ is the smoothing length for discretizing the solid 
and $h^\chi = \max\left(h^s, h^k \right)$ with $k = l, h$ is determined by the corresponding fluid phase interacting with solid. 
Note that subscripts $a$ and $b$ are introduced to denote solid particles and the gradient of the kernel function
\begin{equation}\label{strongkernel}
\nabla_a^0 W_{ab}^{h^s} = \frac{\partial W\left( \mathbf r_{ab}^0, h^s \right)}  {\partial r_{ab}^0} \mathbf e_{ab}^0
\end{equation}
is evaluated at the initial reference configuration.  
Also, the first Piola-Kirchhoff stress tensor is computed from the constitutive law with the deformation tensor $\mathbb{F}$ given by
\begin{equation}
\mathbb{F}_a = \left( \sum_b V^0_b \left( \mathbf{u}_b - \mathbf{u}_a \right) \otimes \nabla^0_a W_{ab}^{h^s}  \right) \mathbb{B}^0_a + \mathbb{I} .
\end{equation}
Following Ref. \cite{zhang2021simple},  
the Kelvin–Voigt type artificial damping formulation  
is applied to the total Lagrangian formulation to enhance the robustness and accuracy.   
\subsection{Treatment of fluid-structure interface}\label{sec:boundary}
\begin{figure}[htb!]
	\centering
	\includegraphics[trim = 2mm 2mm 1mm 2mm, clip, width= 0.45\textwidth]{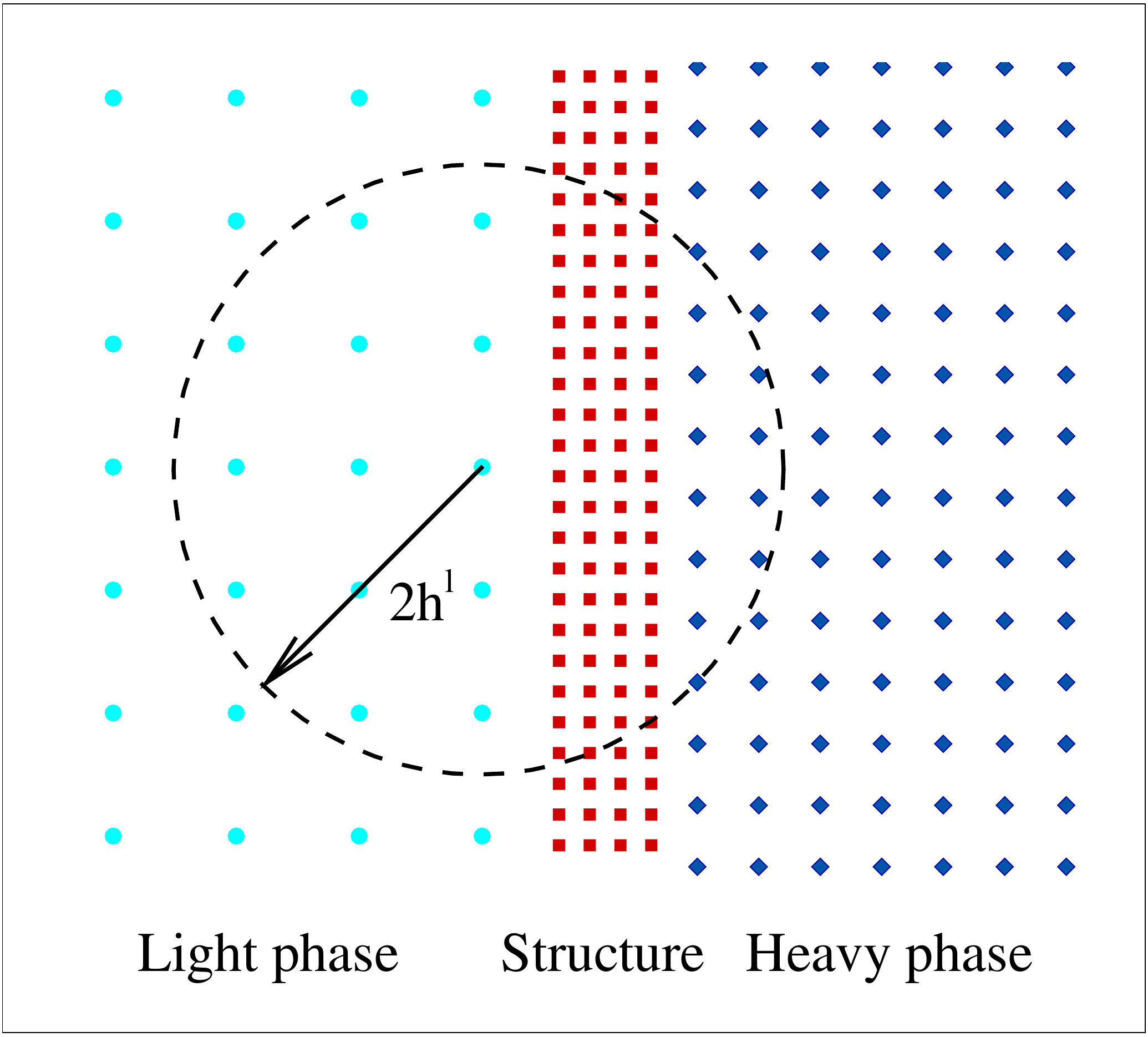}
	\includegraphics[trim = 2mm 2mm 1mm 2mm, clip, width= 0.45\textwidth]{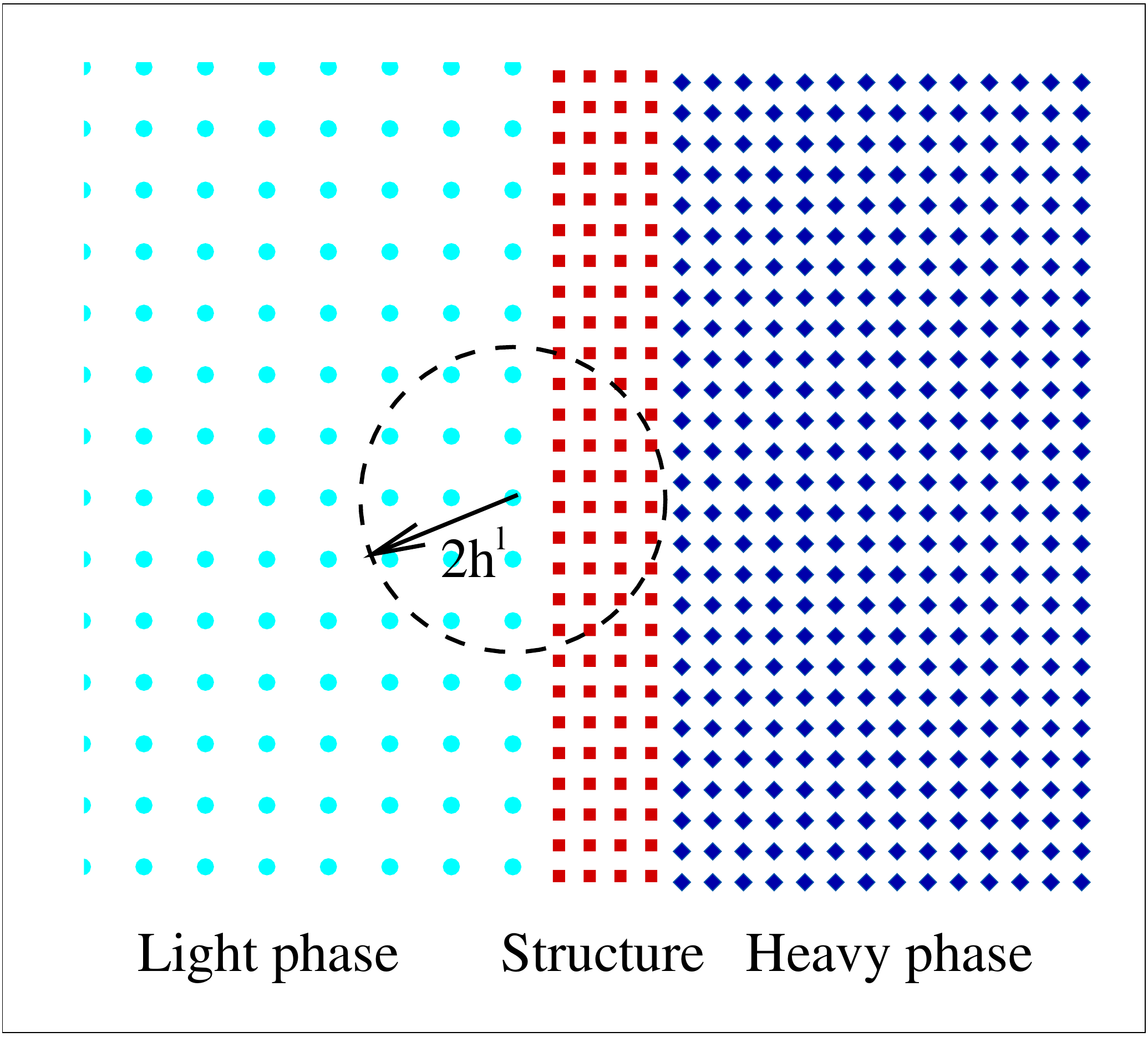}
	\caption{Schematic illustration of the multi-phase FSI discretization in the present multi-resolution framework: 
		Three-level discretization with $h^l = 2.0 h^h = 4.0 h^s$ (left panel) and two-level discretization $h^l = 2.0 h^h = 2.0 h^s$ (right panel).}
	\label{figs:fsi-mr-schematic}
\end{figure}
In Section \ref{sec:fluid-discretization}, 
we have presented the implementation of the multi-resolution framework \cite{zhang2021multi} to address the mutli-resolution discretization of 
multi-phase flow. 
When implementing this framework to solve multi-phase FSI problems, 
one can apply three resolutions for different sub-systems, 
for example three-level discretization with $h^l = 2.0 h^h = 4.0 h^s$,
as shown in the left panel of Figure \ref{figs:fsi-mr-schematic}. 
In this case, 
the computation efficiency can be maximally improved, 
while two drawbacks are induced. 
One is that three-level discretization leads to a smoothing ratio of $h^l / h^s = 4$ between the light phase and structure 
which does not change the numerical accuracy of the FSI coupling while leads to less regular structure oscillations as demonstrated in Ref. \cite{zhang2021multi}. 
Another is that the minimum particle resolution to discretize the structure should be properly adopted to avoid the unphysical communications 
between fluid particles and another ones located on the other side of the structure, 
as shown in Figure \ref{figs:fsi-mr-schematic} (left panel) 
where $4$ layers of particles is applied to discretize the structure and the cut-off radius of $2.6 dp$ is applied. 
Otherwise extra numerical treatment which is not the subjective of this paper should be implemented to avoid this unphysical communications as Ref.~\cite{MCLOONE2022117}. 
To avoid these drawbacks with the achievement of moderate improvement of computational efficiency,  
we apply  two-level discretization with $h^l = 2.0 h^h = 2.0 h^s$, 
as shown in Figure \ref{figs:fsi-mr-schematic} (right panel), 
implying a multi-resolution discretization in multi-phase and light-phase-structure coupling and single-resolution one in heavy-phase-structure coupling. 

\begin{figure}[htb!]
	\centering
	\includegraphics[trim = 3cm 6cm 2cm 4cm, clip, width=.95\textwidth]{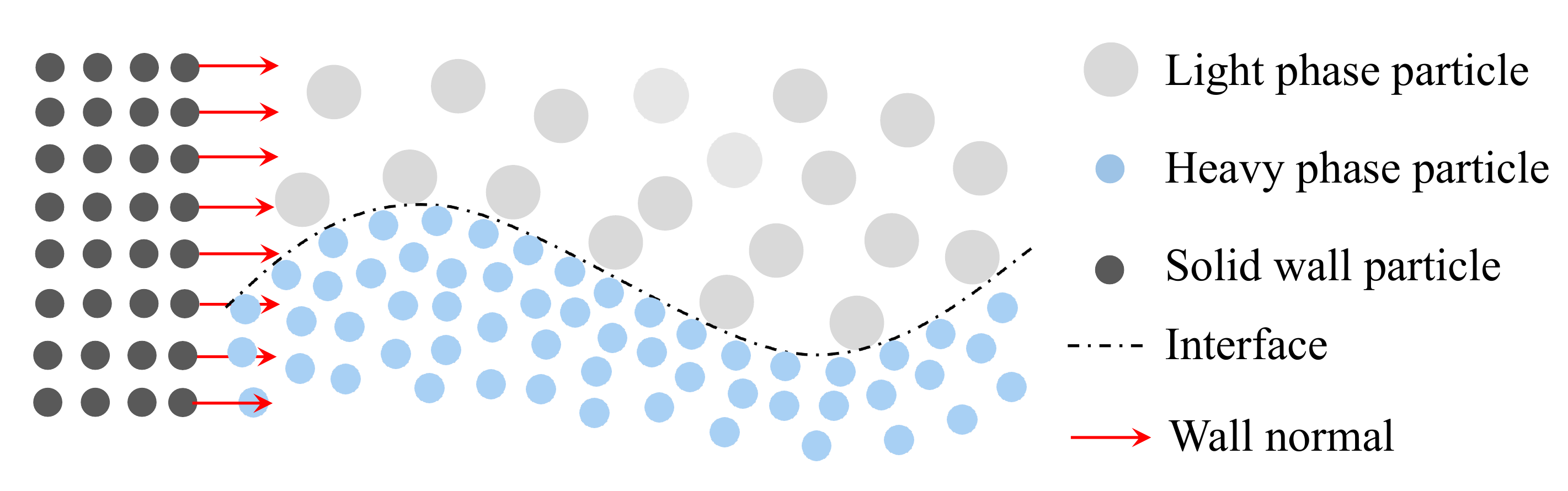}
	\caption{Sketch of multi-phase particles interacting with solid particles along the normal direction through the one-side Riemann based scheme.}
	\label{figs:fsi-wall}
\end{figure}
In the present method, 
the fluid-structure, 
i.e., being rigid or flexible, 
coupling is resolved by the generalized solid boundary condition proposed by Zhang \etal \cite{zhang2022} 
where the flexible structure is behaving as a moving solid boundary for fluid 
and the solid boundary condition is imposed by exploiting a one-sided Riemann problem along the solid-normal direction as shown in Figure \ref{figs:fsi-wall}. 
Following Refs. \cite{zhang2017weakly, zhang2022}, 
the pressure and viscous forces $\mathbf{f}_i^{s:p}$ and $\mathbf{f}_i^{s:\nu}$ of Eq. \eqref{eq:rieammsph-mr} acting on a fluid particle $i$ 
which can be light or heavy phase particle, 
due to the presence of the neighboring rigid or flexible solid particle $a$, 
are calculated from 
\begin{equation} \label{eq:fluid-structure-forces}
\begin{cases}
\mathbf f_i^{s:p} \left(h^\chi \right)  = - \frac{2}{m_i}\sum_a V_i V_a p^\ast \nabla_i W_{ia}^{h^\chi}  \\
\mathbf f_i^{s:\nu} \left(h^\chi \right) = \frac{2}{m_i}\sum_a V_i V_a \frac{2\eta_i \eta_j}{\eta_i + \eta_j} \frac{\mathbf v_i - \mathbf v^d_a}{r_{ia}} \frac{\partial W_{ia}^{h^\chi}}{\partial r_{ia}} .
\end{cases}
\end{equation}
Here, 
$h^\chi = h^l $ and $h^\chi = h^h = h^s$  for light- and heavy-phase interacting with solid structure, respectively, 
with the assumption of $h^l = 2.0 h^h = 2.0 h^s$. 
In this case, 
the light-phase-solid and multi-phase interactions are resolved in the multi-resolution scenario 
and the heavy-phase-solid coupling is discretized in the single-resolution one, 
implying both single- and multi-resolution couplings are simultaneously resolved 
in the triple point where light phase, heavy phase and solid particles meet as shown in Figure \ref{figs:fsi-wall}. 
For the one-sided Riemann problem, 
the left and right states are defined as \cite{zhang2017weakly, zhang2022}
\begin{equation}\label{eq:onesided-rie}
\begin{cases}
\left( \rho_L, U_L, p_L\right) = \left( \rho_f,- \mathbf n_a \cdot \mathbf v_i, p_i \right)  \\
\left( \rho_R, U_R, p_R\right) = \left(\rho^d_a, - \mathbf n_a \cdot \mathbf v^d_a, p^d_a \right), 
\end{cases}
\end{equation}
where $\mathbf n_a$ is the local normal vector pointing from solid to fluid, 
$\rho^d_a$ and $p^d_a$ are imaginary density and pressure, respectively. 
For non-slip boundary condition, 
the imaginary pressure $p^d_a$ and velocity $\mathbf v^d_a$ in \eqref{eq:onesided-rie}  can be derived as
\begin{equation} \label{fs-coupling-mr }
\begin{cases}
p_a^d = p_i + \rho_i max(0, (\mathbf{g} - \widetilde{\frac{\text{d} \mathbf{v}_a}{\text{d}t}}) \cdot \mathbf n_a) (\mathbf{r}_{ia} \cdot \mathbf n_a) \\
\mathbf{v}_a^d = 2 \mathbf{v}_i  - \widetilde{\mathbf{v}}_a ,
\end{cases} 
\end{equation}
to address the force-calculation mismatch in multi-resolution scenario \cite{zhang2021multi} by 
introducing the time averaged velocity $\widetilde{\mathbf{v}}_a$ and acceleration $\widetilde{\frac{d\mathbf{v}_a}{dt}}$ 
of solid particles over one fluid acoustic time step \cite{zhang2020dual}. 
With the imaginary pressure $p_a^d$ in hand, 
and the imaginary density $\rho_a^d$ can be calculated through the EoS presented in Eq. \eqref{eq:EoS}. 
Accordingly, 
the fluid forces exerting on the solid structure $\mathbf{f}_a^{f:p} $ and $\mathbf{f}_a^{f:\nu}$ can be obtained straightforwardly. 
Also, 
the one-sided Riemann solver is also applied to the mass conservation equation 
in fluid-solid coupling. 
In the present FSI treatment, 
the influence of the solid to the fluid is resolved by a local particle-particle interaction pattern through 
the one-sided Riemann solver to compute the boundary flux for imposing proper boundary condition, i.e., no-slid solid boundary condition. 
Subsequently, 
it's straightforward to handle the FSI coupling in the triple point where single- and multi-resolution discretizations are co-exist. 

Similar with Ref.~\cite{zhang2022}, 
particle penetration is observed for light phase in the numerical experiments  
when light phase particles are entrapped during heavy phase impacting on the solid boundary. 
To address this issue, 
a penalty force between light-phase-solid interaction to repel the light-phase particles from the solid is adopted as~\cite{zhang2022} 
\begin{equation}\label{eq:penaltyforce}
	\mathbf f_i = -\frac{2}{m_i} \sum_j V_i V_a \Gamma\left(\mathbf r_i, \mathbf r_a  \right)  \mathbf n_a \frac{\partial W(\mathbf r_{ia}, h^\chi )}{\partial r_{ia}} ,
\end{equation}
where the function $\Gamma\left(\mathbf r_i, \mathbf r_a  \right)$ is defined as
\begin{equation}\label{eq:penaltyfunction}
	\Gamma\left(\mathbf r_i, \mathbf r_a  \right) = \left|p_i \left( \mathbf r_{ia} \cdot \mathbf n_a\right)  \right|  
	\begin{cases}
		\left(1.0 - \beta \right)^2 \frac{0.01 h^s}{h^\chi}  & \beta \leq 1.0  \\ 
		0 & \beta > 1.0 
	\end{cases}.
\end{equation}
Here, 
$h^\chi = h^l$ as the penalty force is only imposed on the light phase particles which are close to solid boundary 
and  the parameter $\beta$ is given by
\begin{equation}
	\beta = 2.0 \frac{\mathbf r_{ia} \cdot \mathbf n_a}{h^s}, 
\end{equation}
allowing a monotonically increasing penalty strength. 
Note that the imposing condition of parameter $\beta \leq 1.0$ implies that there is no penalty force imposed 
for particle $2.0 h^s$ faraway from the wall in its normal direction~\cite{zhang2022}. 
Also, 
there is no opposite force imposed on solid particles. 
\subsection{Time integration}\label{sec:integration} 
To further optimize the computational efficiency, 
we apply the dual-criteria time-stepping method \cite{zhang2020dual}, 
where the advection criterion controls the updating of the particle-neighbor list 
and the corresponding computation of the kernel values and gradients 
and the acoustic criterion determines the time integration of the physical variables, 
for the time integration of fluids.  
Following Ref. \cite{zhang2020dual}, 
the advection criterion $\Delta t_{ad}$ and the acoustic criterion $\Delta t_{ac}$ are defined as 
\begin{equation} 
\begin{cases}\label{eq:timestep-size}
\Delta t_{ad} = \min\left(\Delta t_{ad}^l, \Delta t_{ad}^h\right) \\
\Delta t_{ac} = \min\left(\Delta t_{ac}^l, \Delta t_{ac}^h\right)
\end{cases}, 
\end{equation}
with the advection and acoustic criteria defined as 
\begin{equation} 
\begin{cases}\label{eq:timesteps}
\Delta t_{ad}^i   =  0.25 \min\left(\frac{h^i}{U_{max}}, \frac{{h^i}^2}{\eta} \right) \\
\Delta t_{ac}^i   = 0.6 \min \left( \frac{h^i}{c + U_{max}} \right)
\end{cases} i = l,h .
\end{equation}
Note that we apply single-time stepping for multi-phase coupling instead of multi-time stepping 
which increases complexity due to the requirement of surface particle detection of heavy phase 
and the corresponding operation of time average of physical states to enforce momentum conservation in the interface. 
In this case, 
moderate computational efficiency is achieved with the reward of not increasing the complexity of the present method. 

At the beginning of the advection step, 
the fluid density field is reinitialized by Eq. \eqref{eq:rhosumsurface} or \eqref{eq:rhosum}, 
the viscous force is computed and the transport-velocity formulation of Eq. \eqref{eq:transprort} is only applied for the light phase. 
During the advection criterion $\Delta t_{ad}$ , 
the pressure relaxation process is conducted several acoustic time steps with the criterion $\Delta t_{ac}$ 
by using the position-based Verlet scheme \cite{zhang2021multi}.
For each acoustic time step, 
the density and position of fluids are first updated to the mid-point as
\begin{equation}\label{eq:verlet-first-half}
\begin{cases}
\rho_i^{n + \frac{1}{2}} = \rho_i^n + \frac{1}{2}\Delta t_{ac}  \frac{d \rho_i}{dt}\\
\mathbf r_i^{n + \frac{1}{2}} = \mathbf r_i^n + \frac{1}{2} \Delta t_{ac} {\mathbf v_i}^{n} .
\end{cases} 
\end{equation}
Then, 
the velocity of fluids is integrated to the new time step with
\begin{equation}\label{eq:verlet-first-mediate}
\mathbf v_i^{n + 1} = \mathbf v_i^n +  \Delta t_{ac}  \frac{d \mathbf v_i}{dt}. 
\end{equation}
Finally, position and density of fluid are updated another half step as
\begin{equation}\label{eq:verlet-first-final}
\begin{cases}
\mathbf r_i^{n + 1} = \mathbf r_i^ {n + \frac{1}{2}} +  \frac{1}{2} \Delta t_{ac} {\mathbf v_i^{n + 1}} \\
\rho_i^{n + 1} = \rho_i^{n + \frac{1}{2}} + \frac{1}{2} \Delta t_{ac} \frac{d \rho_i}{dt} .
\end{cases} 
\end{equation}
At this point, 
one time step integration of fluids is completed and 
the pressure and viscous forces inducing acceleration to solid due to the existence of fluids 
are computed with Eq. \eqref{eq:fluid-structure-forces} 
and will be considered as constant during the time integration of solid with the criterion  
\begin{equation}\label{eq:solidadvection}
\Delta t^s   =  0.6 \min\left(\frac{h^s}{c^s + |\mathbf{v}|_{max}},
\sqrt{\frac{h^s}{|\frac{\text{d}\mathbf{v}}{\text{d}t}|_{max}}} \right) . 
\end{equation}
where $c^s $ is the artificial speed of sound of flexible solid. 
Following Ref. \cite{zhang2021multi}, 
a multi-time stepping method is applied for the time integration of FSI coupling, implying  
$\kappa = [\frac{\Delta t_{ac}}{\Delta t^s}] + 1$ steps are conducted for solid 
during each acoustic  time step size $\Delta t_{ac}$ of fluids. 
We introduce herein index $\varkappa = 0, 1, ...,  \kappa-1 $ to denote integration step for solid.  
For each time step,  
the deformation tensor, density and position of solid are first updated to the midpoint as 
\begin{equation}\label{verlet-first-half-solid}
\begin{cases}
\mathbb{F}_a^{\varkappa + \frac{1}{2}} = \mathbb{F}_a^{\varkappa} + \frac{1}{2} \Delta t^s \frac{\text{d} \mathbb{F}_a}{\text{d}t}\\
\rho_a^{\varkappa + \frac{1}{2}} = \rho_a^0 \frac{1}{J} \\
\mathbf{r}_a^{\varkappa + \frac{1}{2}} = \mathbf{r}_a^{\varkappa} + \frac{1}{2} \Delta t^s {\mathbf{v}_a} ,
\end{cases} 
\end{equation}
followed by the update of the velocity to the new time step with 
\begin{equation}\label{verlet-first-mediate-solid}
\mathbf{v}_a^{\varkappa + 1} = \mathbf{v}_a^{\varkappa} +  \Delta t^s  \frac{d \mathbf{v}_a}{dt}. 
\end{equation}
Finally, 
the deformation tensor, density and position of solid are updated to the new time step by 
\begin{equation}\label{verlet-first-final-solid}
\begin{cases}
\mathbb{F}_a^{\varkappa + 1} = \mathbb{F}_a^{\varkappa + \frac{1}{2}} + \frac{1}{2} \Delta t^s \frac{\text{d} \mathbb{F}_a}{\text{d}t}\\
\rho_a^{\varkappa + 1} = \rho_a^0 \frac{1}{J} \\
\mathbf{r}_a^{\varkappa + 1} = \mathbf{r}_a^{\varkappa + \frac{1}{2}} + \frac{1}{2} \Delta t^s {\mathbf{v}_a}^{\varkappa + 1}.
\end{cases} 
\end{equation}
It is worth noting that the rate of change for deformation gradient $\frac{\text{d}\mathbb{F}}{\text{d}t}$ is computed through \cite{zhang2021integrative} 
\begin{equation}\label{defmoration-reate}
\frac{\text{d}\mathbb{F}_a}{\text{d}t}  = \left( \sum_b V^0_b \left( \mathbf{v}_b - \mathbf{v}_a \right) \otimes \nabla^0_a W_{ab}  \right) \mathbb{B}^0_a. 
\end{equation}
%
%
%
\clearpage
\section{Numerical examples}\label{sec:examples}
In this section, 
several numerical examples of two-phase flow involving high-density ratio, complex phase change and interaction with rigid or flexible solid 
are investigated to assess the robustness, accuracy and efficiency of the present method by qualitative and quantitative comparison with 
experimental and numerical data in literature. 
In all tests, 
the 5th-order Wendland kernel \cite{Wendland1995} where 
smoothing length $h = 1.3 dp$ and support radius of $2h$
with $dp$ denoting the initial particle spacing are applied. 
To setup the artificial speed of sound $c^f$, 
the maximum anticipated flow speed  is estimated as $U_{max} = 2\sqrt{gH}$, 
where $H$ is the initial water depth, 
following to the shallow-water theory \cite{ritter1892fortpflanzung} 
without special specification. 
\subsection{Hydrostatic test }\label{sec:hydrostatic}
\begin{figure}[htb!]
	\centering
	\includegraphics[trim = 10cm 4.5cm 10cm 2cm, clip, width= 0.5\textwidth]{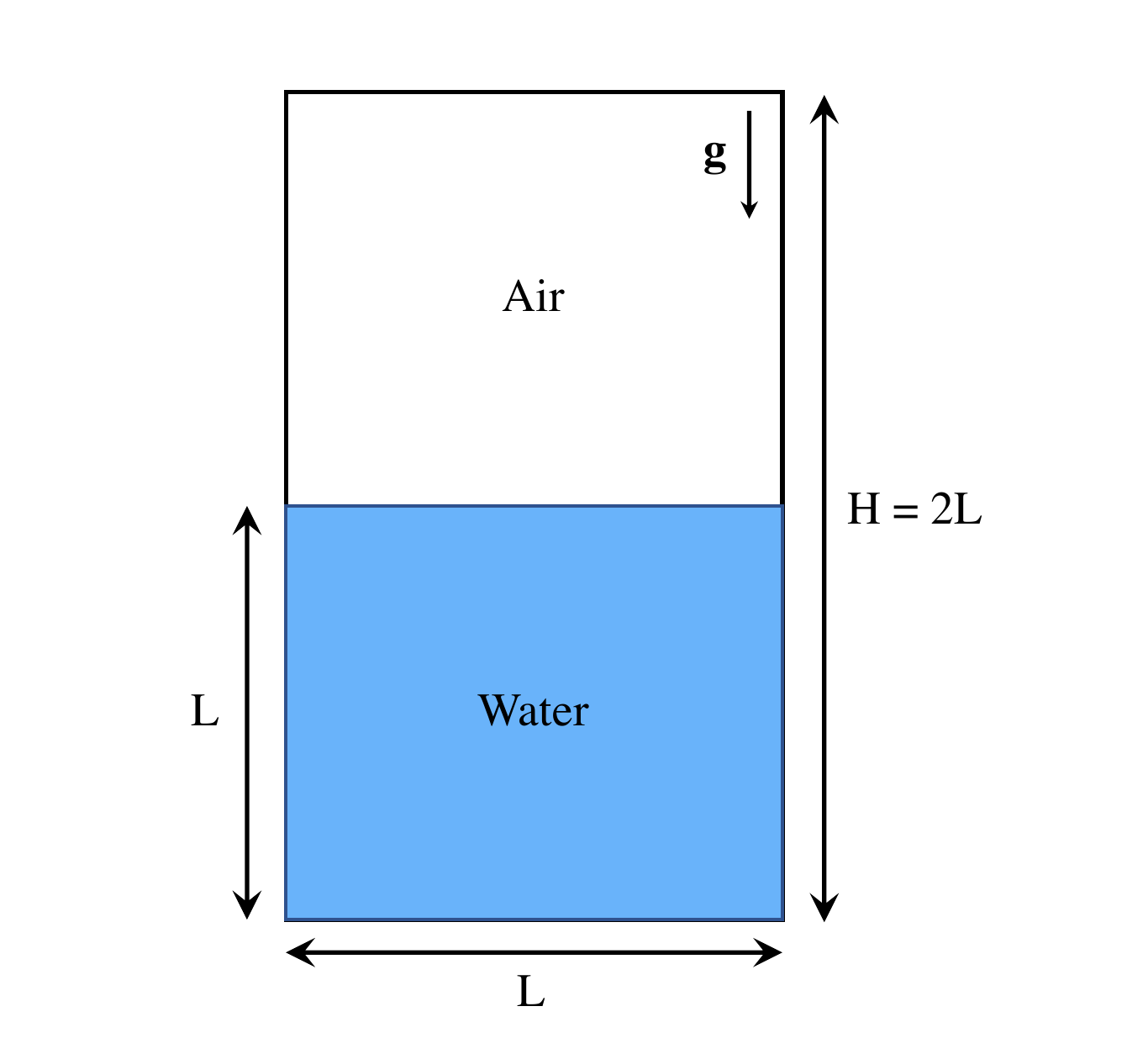}
	\caption{Hydrostatic test: Schematic illustration.}
	\label{figs:hydrostatic-setup}
\end{figure}
The first benchmark test we considered herein is a two-dimensional two-phase hydrostatic test 
in which gravity presents as shown in Figure \ref{figs:hydrostatic-setup} 
to assess the compatibility with the static solution and the ability of capturing sharp interface under high density ratio 
of the present method.  
Following Ref. \cite{rezavand2020weakly}, 
the tank has a length of $L = 1$ and height of $H = 2L$ with the lower and upper half parts are filled with water and air, respectively. 
Both fluids are initially at rest and considered to be inviscid. 
We set water and air density as $\rho_w = 1$ and $\rho_a = 0.001$, respectively, 
resulting a density ratio of $1000$. 
Note that unit gravity acceleration in negative $y$-direction is applied and all other quantities correspond to their non-dimensional variables. 
In present simulation, 
the particles are placed on a Cartesian lattice with a particle spacing of $dp^w = L/50$ for discretizing the water. 
For single- and multi-resolution simulations, 
we consider the water-air two cases with resolution ratio ${dp}^a/{dp}^w = 1$ and ${dp}^a/{dp}^w = 2$ 
to investigate the computational efficiency and accuracy. 

\begin{figure}[htb!]
	\centering
	\includegraphics[trim = 5mm 1mm 5mm 1mm, clip, width=\textwidth]{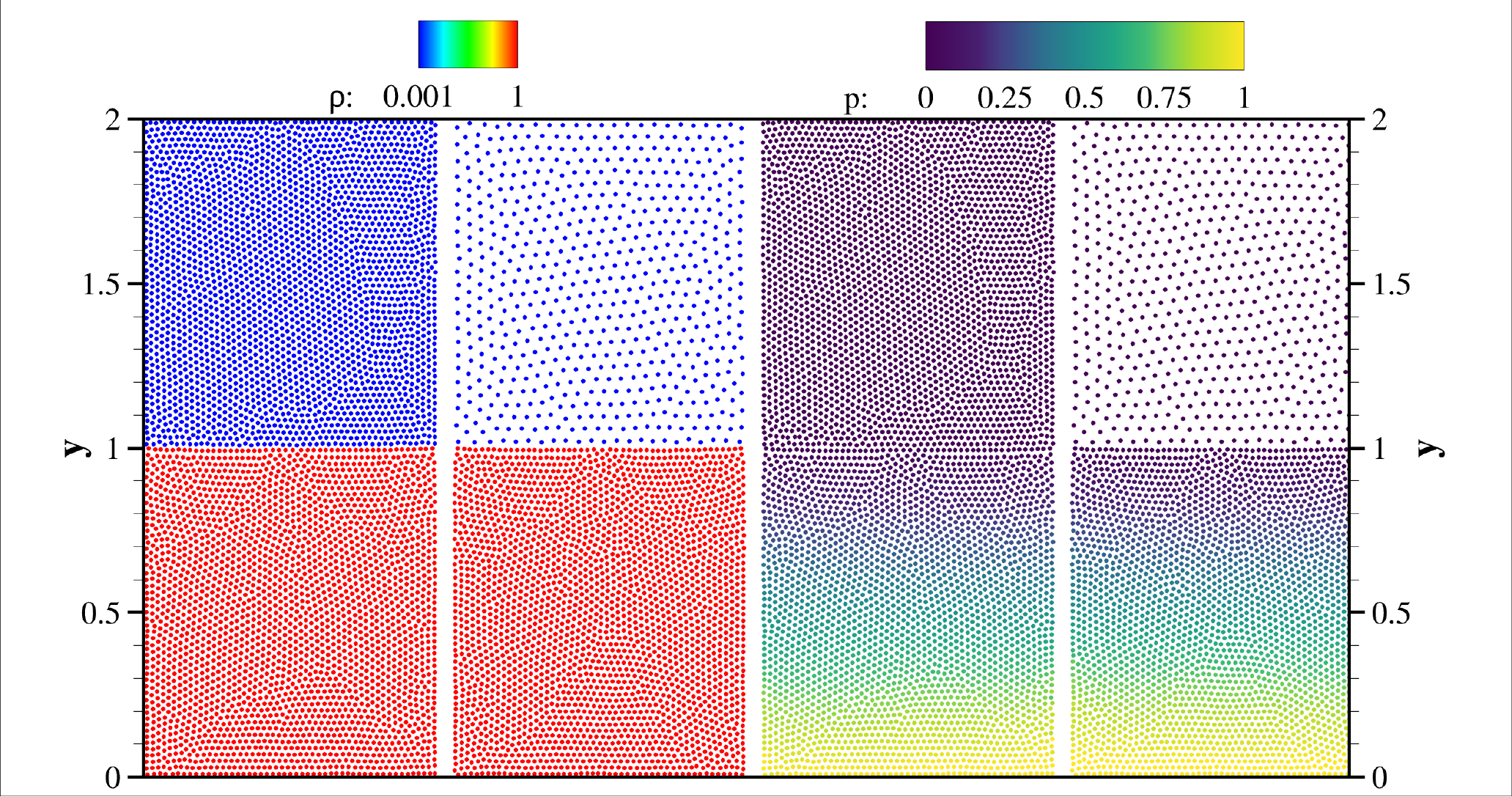}
	\caption{
		Hydrostatic test: 
		Particle distributions with density (left panel) and pressure contours (right panel) in single- and multi-resolution scenarios. 
		(For interpretation of the references to color in this figure legend, the reader is referred to the web version of this article.)
	}
	\label{figs:hydrostatic}
\end{figure}
Figure \ref{figs:hydrostatic} portrays the particle distribution at $t = 50$ with density (left panel) and pressure (right panel) contours
of the hydrostatic test in single- and multi-resolution scenarios. 
It can be evidently observed that a sharp interface without any notable non-physical motion of water surface
and a smooth pressure filed are obtained in both single- and multi-resolution simulations. 
These observations demonstrate the accuracy and robustness of the present method in capturing sharp interface 
and being compatible with hydrostatic solution without exhibiting unphysical surface motion.  

\begin{figure}[htb!]
	\centering
	\includegraphics[trim = 2mm 2mm 2mm 2mm, clip, width=\textwidth]{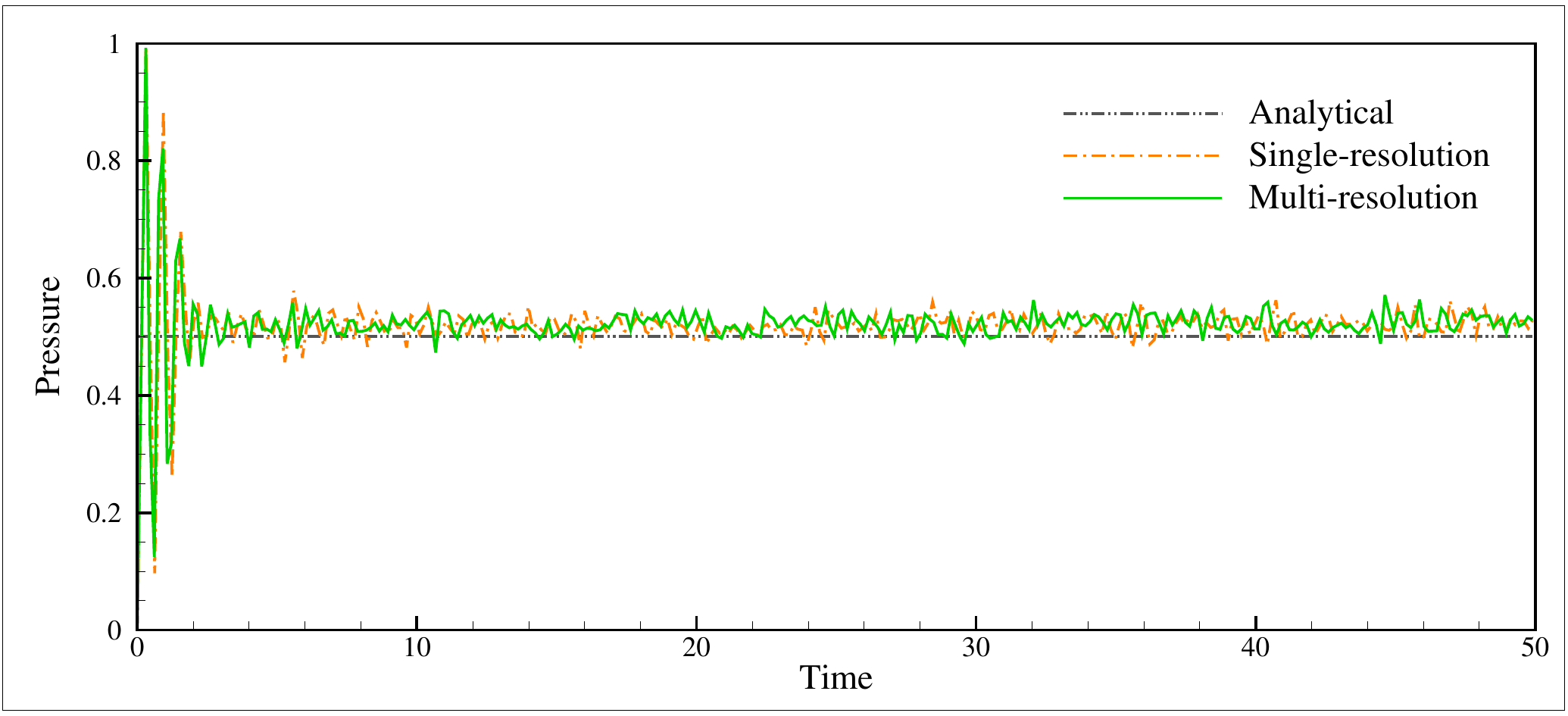}
	\caption{
		Hydrostatic test: 
		Time histories of the numerical and analytical pressures probed at $y = 0.25 H$ in single- and multi-resolution scenarios.
		(For interpretation of the references to color in this figure legend, the reader is referred to the web version of this article.)	
	}
	\label{figs:hydrostatic-pressure}
\end{figure}
\begin{figure}[htb!]
	\centering
	\includegraphics[trim = 2mm 2mm 2mm 2mm, clip, width=\textwidth]{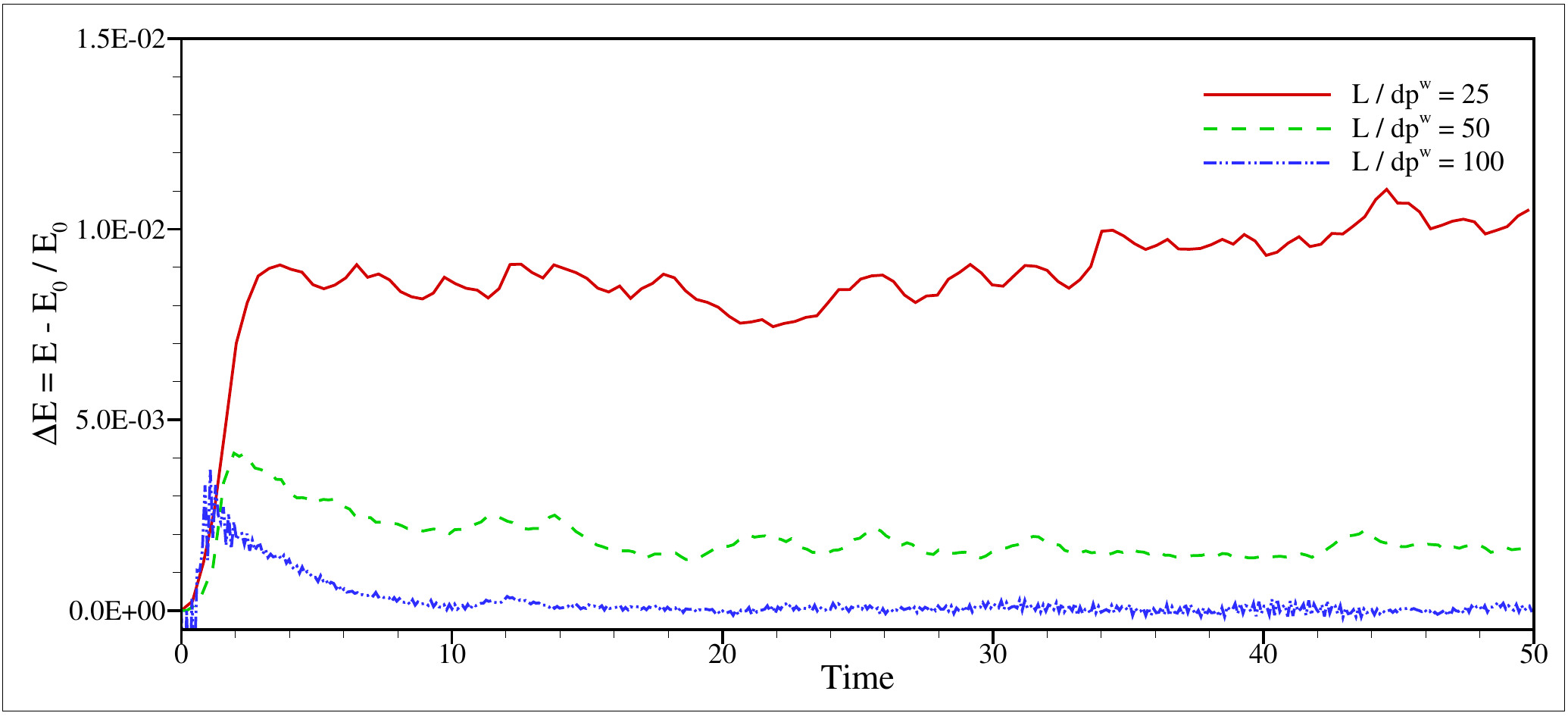}
	\caption{
		Hydrostatic test: 
		Time evolutions of the normalized mechanical energy in multi-resolution simulation.
		(For interpretation of the references to color in this figure legend, the reader is referred to the web version of this article.)
	}
	\label{figs:hydrostatic-energy}
\end{figure}

The accuracy of the present method is further quantitatively assessed by comparing 
the numerically computed pressure profile with the analytical hydrostatic pressure value. 
Figure \ref{figs:hydrostatic-pressure} presents the time history of the numerical and analytical pressure signals probed at $y = 0.25 H$. 
For both single- and multi-resolution computations, 
the numerical pressure file has a good agreement with the analytical values, demonstrating the accuracy of the present method. 
Furthermore, 
the energy conservation property is also investigated to show the substantial characteristic of the present method. 
Figure \ref{figs:hydrostatic-energy} plots the time evolution of the normalized mechanical energy 
defined as $\frac{E - E_0}{E_0}$ with $E_0$ denoting the initial value 
for single- and multi-resolution simulations.  
It can be observed that both the normalized mechanical energies rapidly decay to a very small value 
after an early-stage oscillations induced by the weakly-compressible model.  

In general, 
the present method is compatible with hydrostatic solution and able to capture the sharp interface with high density ratio of $1000$. 
Also, 
the multi-resolution method shows identical numerical accuracy 
in comparison with single-resolution counterpart while achieves a speedup of $1.63$ as shown in Table \ref{tab:hydrostatic-fsi-cputime}.
\begin{table}[htb!]
	\centering
	\caption{Hydrostatic test: Analysis of computational efficiency. 
		The computations are carried out on an Intel(R) Xeon(R) L5520 2.27GHz desktop 
		computer with 24GiB RAM and a Scientific Linux operating system (6.9). 
		To analyze the computational performance,  
		we evaluate the CPU wall-clock time of $L/dp^w = 100$ for shared memory parallelized computations (based on the Intel TBB library) until $50$ time instant. }
	\begin{tabular}{ccccc}
		\hline
		Cases  & Single resolution & Multi resolution & Speedup \\
		\hline
		CPU time (s)	&   741.7 & 453.71 & 1.63     \\ 
		\hline
	\end{tabular}
	\label{tab:hydrostatic-fsi-cputime}
\end{table}
%
\subsection{Dam-break flow}\label{sec:dambreak}
\begin{figure}[htb!]
	\centering
	\includegraphics[width=0.75\textwidth]{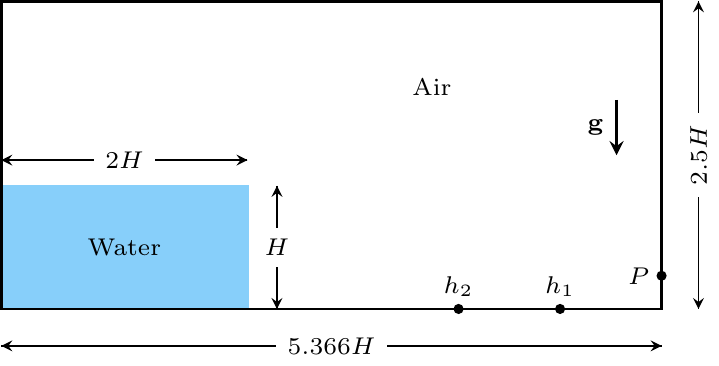}
	\caption{
		Dam-break flow: Schematic illustration. 
		Here, the pressure probe $P$ is located at $ y / H = 0.19$, and two surface sensors $h_1$ and $h_2$ are located 
		$x_2 / H = 0.825$ and $x_2 / H = 1.653$ away from the right wall, respectively. 
	}
	\label{figs:dam-setup}
\end{figure}
In this section, 
we consider a two dimensional two-phase dam-break flow 
which exhibits complex interface change owing to violent wave impact and breaking events. 
As a benchmark test, 
this problem has been comprehensively studied in literature by experiment \cite{buchner2002green, martin1952part, lobovsky2014experimental} 
and numerical modeling \cite{colagrossi2003numerical, rezavand2020weakly, zhang2017weakly, wang2019improved, adami2012generalized} 
with or without considering the air phase. 
Following Refs.\cite{lobovsky2014experimental, rezavand2020weakly, zhang2017weakly}, 
the schematic of this problem is illustrated in Figure \ref{figs:dam-setup}. 
Initially, 
a water column with height of $H=1$ and width of $2H$ is surrounded by air and located 
at the left corner of a tank with length $5.366H$ and height $2.5 H$. 
In the present study, 
the flow is considered to be inviscid and the density of water and air are set to 
$\rho_w = 1$ and $\rho_a = 0.001$, respectively, resulting a density ratio of $1000$. 
With the gravity $g=-1$ in $y$-direction, 
all quantities correspond to their non-dimensional variables.
Similar with Refs. \cite{rezavand2020weakly, zhang2017weakly}, 
the water column with zero initial pressure is released immediately as the computation starts 
other than being released from 
an up-moving gate as in the experiment \cite{buchner2002green, lobovsky2014experimental}, implying pressure-relaxed water.
For quantitative comparison, 
one pressure probe $P$ located at the downstream wall at $y/H=0.19$ \cite{adami2012generalized, rezavand2020weakly} 
and two surface sensors $h_1$ and $h_2$ respectively located at $x_2 / H = 0.825$ and $x_2 / H = 1.653$ away from the right wall 
are set for recording the impact pressure signal and water surface evolution. 
Following Refs. \cite{adami2012generalized, rezavand2020weakly}, 
the probe position is slightly shifted in comparison with the experimental setup \cite{buchner2002green}
as suggested by Greco \cite{greco2001two} to produce a better agreement. 
For multi-resolution simulations, 
we consider the water-air resolution ratio ${dp}^a/{dp}^w = 2$. 

\begin{figure}[htb!]
	\centering
	\includegraphics[trim = 5mm 1mm 5mm 2mm, clip, width=\textwidth]{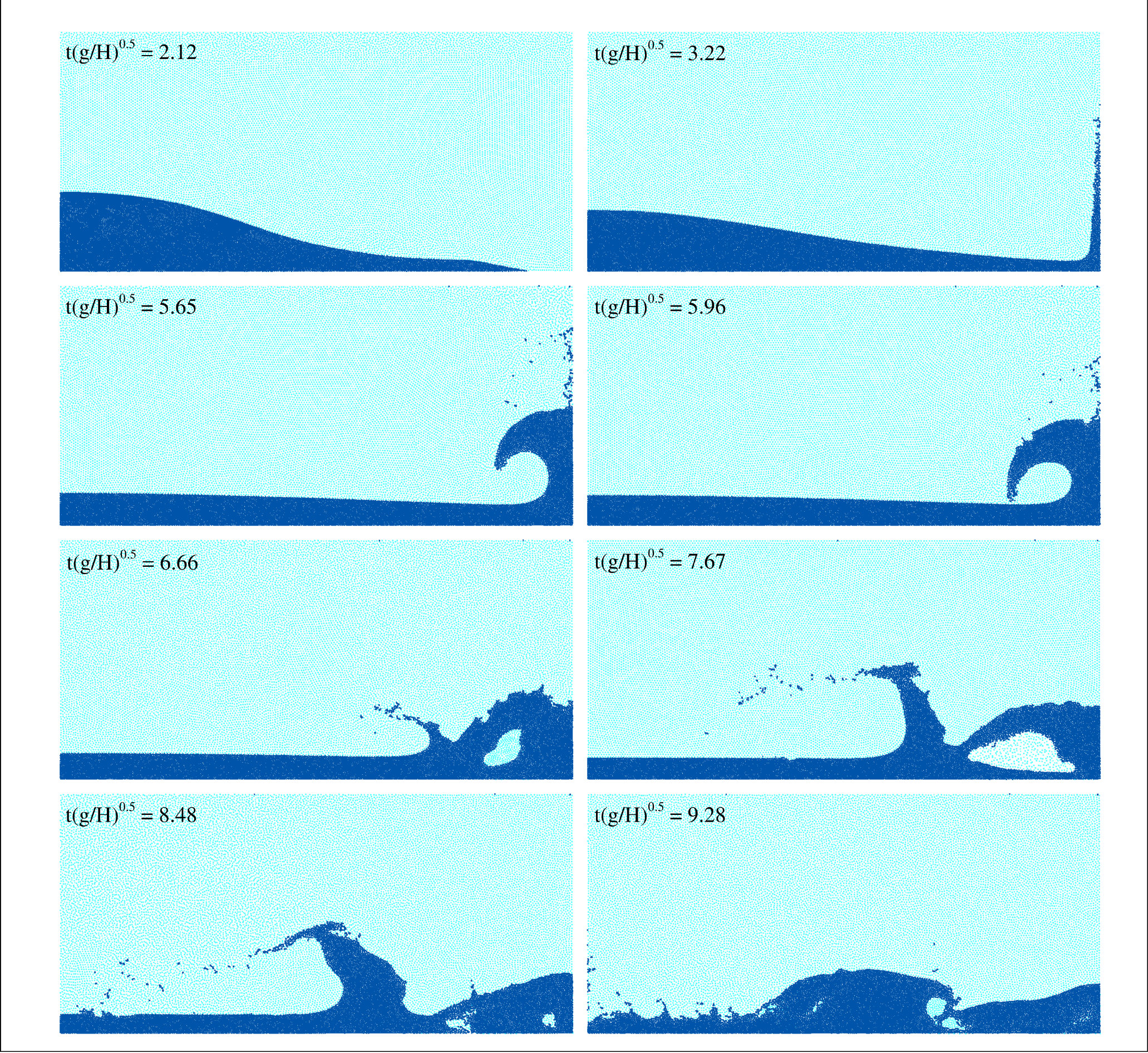}
	\caption{
		Dam-break flow: 
		Snapshots for the evolution of the water-air interface at different time instants obtained by the
		multi-resolution simulation. 
		(For interpretation of the references to color in this figure legend, the reader is referred to the web version of this article.)
	}
	\label{figs:dam-particle}
\end{figure}
Figure \ref{figs:dam-particle} illustrates several snapshots at different time instants for the water-air interface evolution 
obtained by the multi-resolution simulation. 
Similar with previous single-phase \cite{ferrari2009new, adami2012generalized, zhang2017weakly} 
and multi-phase \cite{colagrossi2003numerical, rezavand2020weakly} simulation results, 
the main features of the dam-break flow, 
i.e., 
high roll-up along the downstream wall and the induced first jet due to its falling, 
the re-entry of the first jet and a large secondary reflective jet due to the backward wave motion, 
are well captured by the present multi-resolution simulation.
More importantly, 
the sharp water-air interface is well maintained without exhibiting any unnatural void regions and unrealistic phase separation, 
implying the robustness of the present multi-resolution method. 

\begin{figure}[htb!]
	\centering
	\includegraphics[trim = 1mm 1mm 1mm 1mm, clip, width=\textwidth]{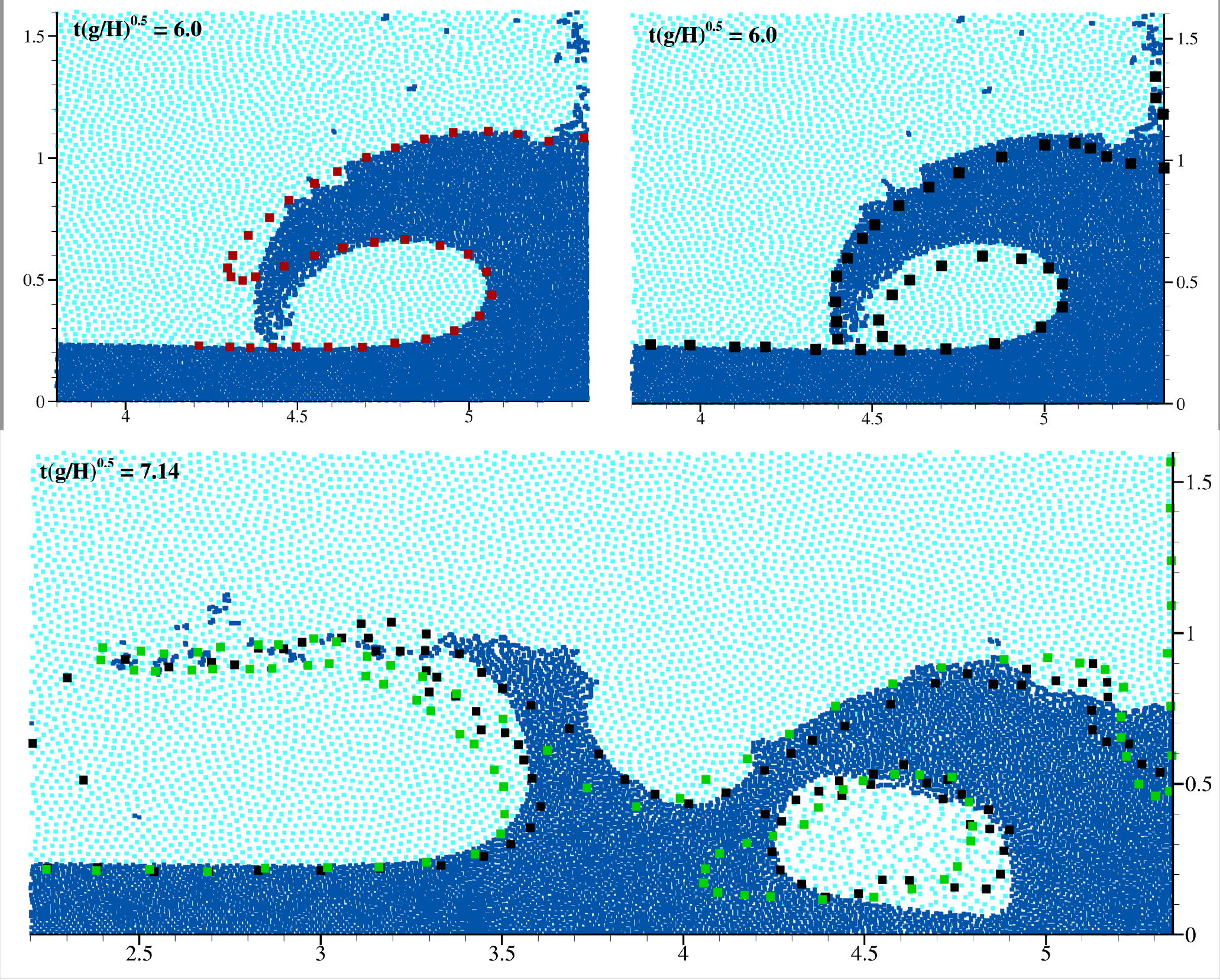}
	\caption{
		Dam-break flow: Evolution of the water-air interface 
		obtained by the present multi-resolution SPH method with $H/dp^w=80$, 
		in comparison with those obtained by the BEM method (\crule[red]{2mm}{2mm}) in Ref. \cite{antuono2012numerical}, 
		the two-phase SPH method (\crule{2mm}{2mm}) in Ref.  \cite{rezavand2020weakly}
		and the two-phase level-set method (\crule[green]{2mm}{2mm}) in Ref. \cite{colicchio2005level}. 
		(For interpretation of the references to color in this figure legend, the reader is referred to the web version of this article.)
	}
	\label{figs:dam-particle-comparison}
\end{figure}
To qualitatively assess the accuracy of the present method in capturing the complex interface evolution, 
the water-air interface profile at time instants of $\text t \left(\mathbf g/ \text H \right)^{0.5} = 6.0 $ and $\text t \left(\mathbf g/ \text H \right)^{0.5} = 7.14$ 
is compared with those obtained by the boundary element method (BEM) \cite{antuono2012numerical}, 
the level-set method \cite{colicchio2005level} and the multi-phase SPH method \cite{rezavand2020weakly}. 
In present multi-resolution simulation with water-air resolution ratio ${dp}^a/{dp}^w = 2$ and $H/dp^w = 80$ which is identical to the one applied in Ref. \cite{rezavand2020weakly}, 
a sharp water-air interface is well captured and maintained at time instant $\text t \left(\mathbf g/ \text H \right)^{0.5} = 6.0 $
before the re-entry of the backward wave 
and the jet due to the falling of the run-up along the right-hand-side wall 
shows a good agreement with those predicted by the BEM \cite{colicchio2005level} and SPH method \cite{rezavand2020weakly}. 
Compared with the SPH prediction, 
the BEM result exhibits excessive numerical dissipation and 
shows considerable time delay of the backward wave, 
as also observed in Refs. \cite{antuono2012numerical, rezavand2020weakly}. 
At the wave breaking stage at $\text t \left(\mathbf g/ \text H \right)^{0.5} = 7.14 $, 
a large secondary reflective jet and an entrapped air cavity are reasonably predicted by the present multi-resolution method 
compared with those obtained by the Level-set and multi-phase SPH methods \cite{colicchio2005level, rezavand2020weakly}, 
while slight discrepancies are noted due to the physical uncertainties induced by complex interface change. 
In general, 
the present multi-resolution method can accurately predict the complex interface, 
meanwhile, 
maintains a sharp phase interface without unnatural void region as in Ref. \cite{rezavand2020weakly} 
and shows less numerical dissipation and improved computational efficiency as discussed in the following part. 

\begin{figure}[htb!]
	\centering
	\includegraphics[trim = 2mm 2mm 2mm 2mm, clip, width=\textwidth]{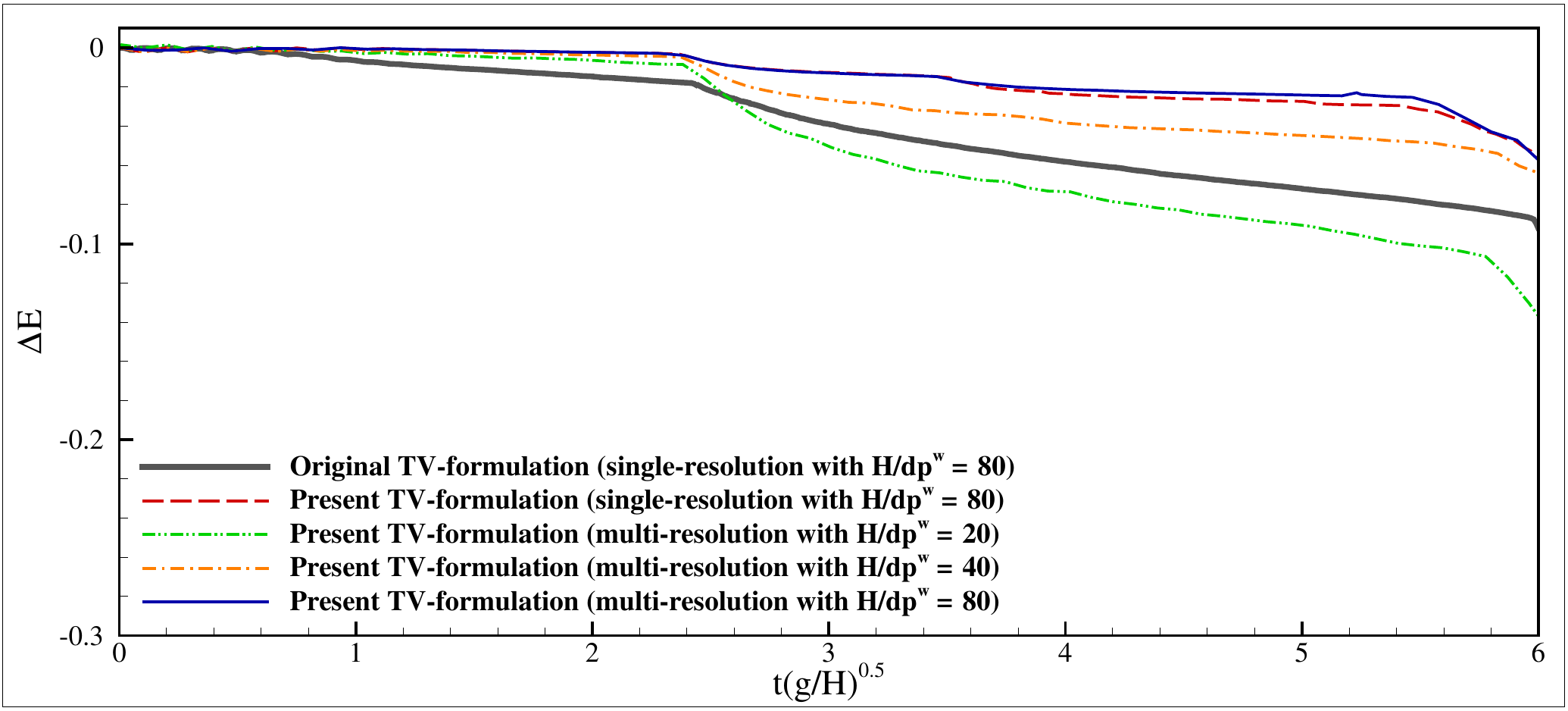}
	\caption{
		Dam-break flow: 
		The time history of the numerical dissipation of mechanical energy by using the present modified 
		transport-velocity formulation with $\widehat{p}_0  = 10 \rho_i U^2_{max}$ with $U_{max}$
		and the original one with $\widehat{p}_0  = 4 \rho^0 {c^f}^2$. 
		Also, a convergence study is conducted for the present formulation. 
		(For interpretation of the references to color in this figure legend, the reader is referred to the web version of this article.)
	}
	\label{figs:dam-energy}
\end{figure}
To demonstrate the low dissipation property of the present modification to the transport-velocity formulation, 
we assess the numerical dissipation of the mechanical energy by defining \cite{zhang2017weakly, rezavand2020weakly, marrone2010fast}
\begin{equation}
		\Delta E = \frac{E_{k}+E_{p}-E_{p}^0}{E_{p}^0-E_{p}^{\infty}}, 
\end{equation}
where 
$E_{k}$ is the kinetic energy, 
$E_{p}$ the potential energy, 
$E_{p}^0$ the initial potential energy and 
$E_{p}^{\infty}$ the potential energy when the flow reaches a hydrostatic state finally.
Figure \ref{figs:dam-energy} plots the time evolution of the 
numerical dissipation of mechanical energy in single- and multi-resolution simulations 
with different background pressures. 
In the single-resolution simulation, 
the present modification of the background pressure evidently exhibits much less numerical dissipation. 
Also, 
identical numerical dissipation is achieved by both single- and multi-resolution simulations with the present modification. 
In addition, 
a convergence study for the multi-resolution method with the background pressure modification is also depicted in Figure\ref{figs:dam-energy}. 
It can be noted that the numerical dissipation is rapidly decreased with increasing the spatial resolution. 

\begin{figure}[htb!]
	\centering
	\includegraphics[trim = 2mm 2mm 2mm 2mm, clip, width=\textwidth]{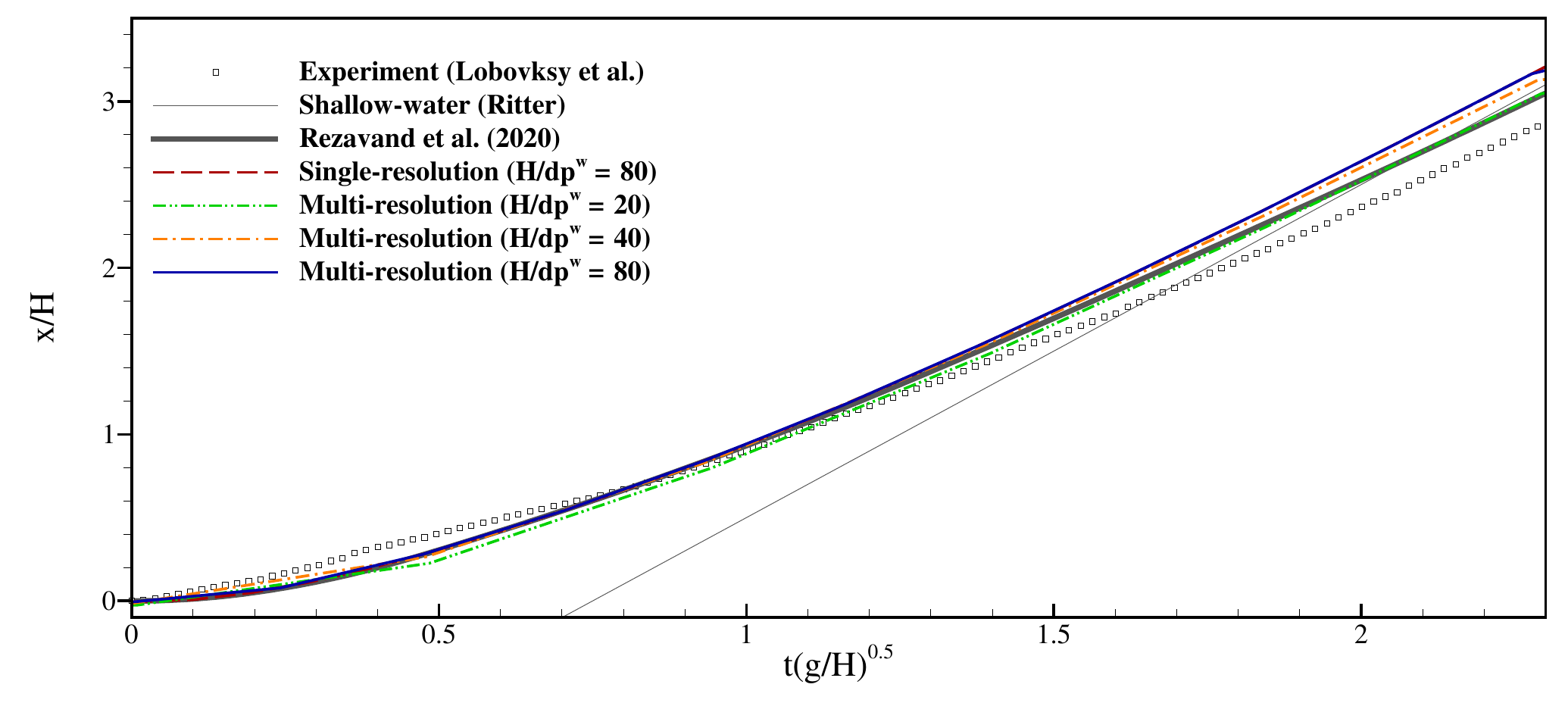}
	\caption{
		Dam-break flow: 
		The time history of the water wave front in single- and multi-resolution simulations in comparison 
		with the analytical solution derived by shallow water theory \cite{ritter1892fortpflanzung}, 
		experimental data \cite{lobovsky2014experimental} and 
		numerical results by using the multi-phase SPH method \cite{rezavand2020weakly}. 
		Note that a convergence study of the present multi-resolution method is also presented herein.
		(For interpretation of the references to color in this figure legend, the reader is referred to the web version of this article.)
	}
	\label{figs:dam-wave-front}
\end{figure}
Figure \ref{figs:dam-wave-front} plots the time history of the predicted water wave front in single- and multi-resolution simulations
and its comparison with analytical solution obtained with shallow water theory \cite{ritter1892fortpflanzung}, 
experimental data \cite{lobovsky2014experimental} and numerical results using multi-phase SPH method \cite{rezavand2020weakly}. 
As expected, 
both single- and multi-resolution results have a good agreement with the theoretical shallow water solution, 
while show slight faster front wave propagation compared with the experimental observation due to the uncertainties of 
repeatability of experiment, wall roughness and the turbulence effects of boundary layer. 
Compared with the results presented in Ref. \cite{rezavand2020weakly}, 
the present results show faster front wave propagation due to the modification of transport-velocity formulation 
and the implementation of solid boundary condition with Riemann-based scheme \cite{zhang2022}. 
A convergence study of the present multi-resolution method is also reported in Figure \ref{figs:dam-wave-front}. 
With the increase of the spatial resolution, 
the numerical predicted propagation speed of the wave front is rapidly converged to 
the analytical solution with shallow water theory. 
Note that the shallow water assumption does not hold at initial time instants of the dam-break flow.

\begin{figure}[htb!]
	\centering
	\includegraphics[trim = 2mm 2mm 2mm 2mm, clip, width=\textwidth]{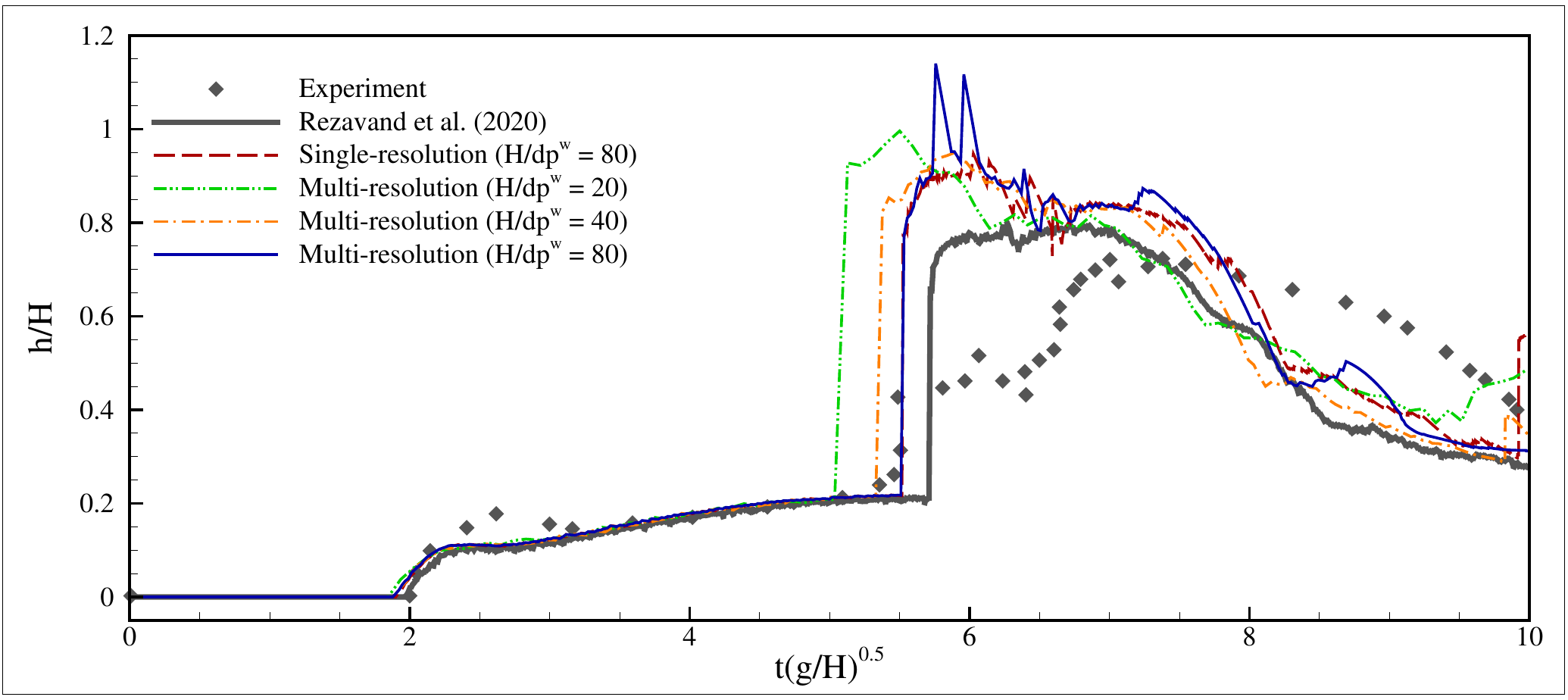}
	\includegraphics[trim = 2mm 2mm 2mm 2mm, clip, width=\textwidth]{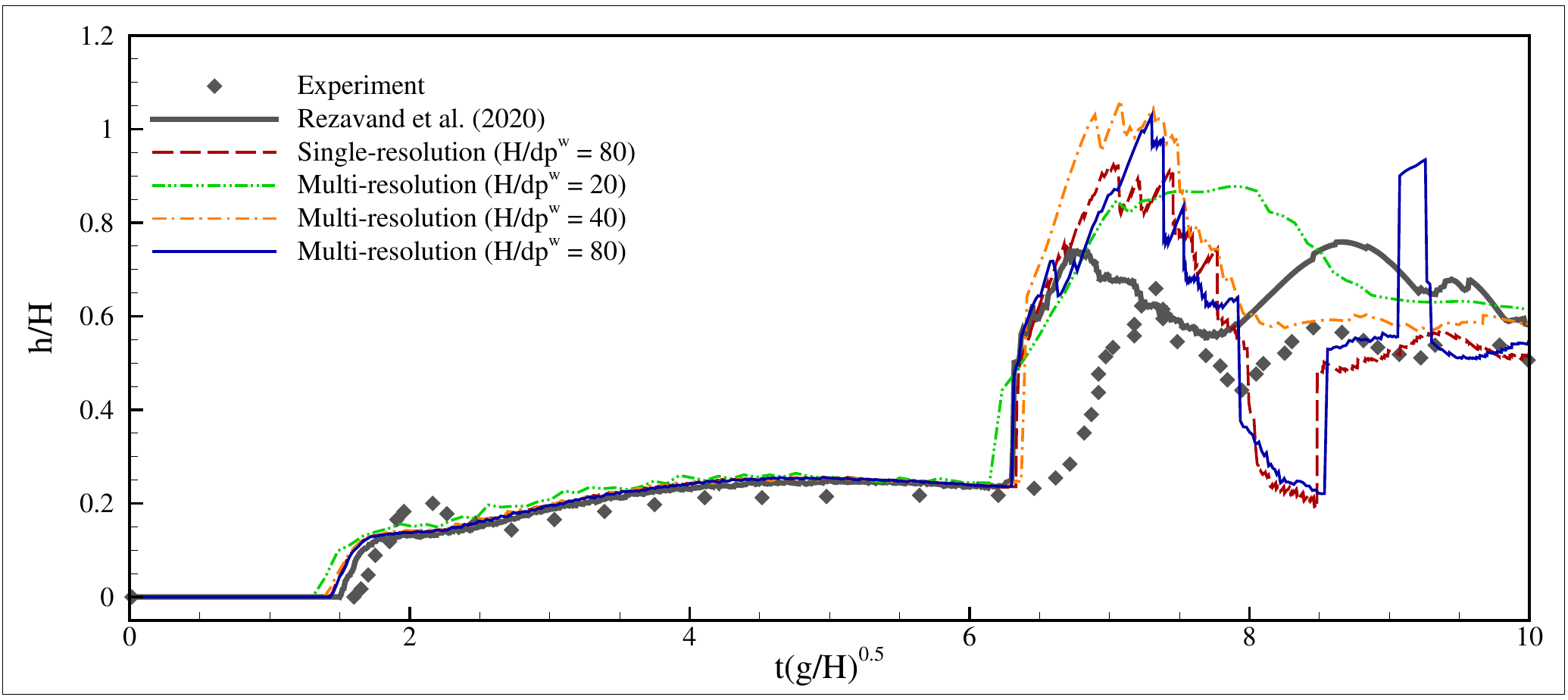}
	\caption{
		Dam-break flow: 
		The time evolution of the water surface probed at $h_1$ (upper panel) and $h_2$ (bottom panel) in single- and multi-resolution simulations
		and its comparison with experimental observation \cite{lobovsky2014experimental} 
		and numerical result obtained by Ref. \cite{rezavand2020weakly}. 
		The convergence study of the multi-resolution method is also presented herein. 
		(For interpretation of the references to color in this figure legend, the reader is referred to the web version of this article.)
	}
	\label{figs:dam-surface-probe}
\end{figure}
Figure \ref{figs:dam-surface-probe} depicts the time history of the water surface measured 
by sensors $h_1$ (upper panel) and $h_2$ (bottom panel) obtained by the present method in single- and multi-resolution scenarios, 
and its comparison with experimental and numerical data \cite{lobovsky2014experimental, rezavand2020weakly}. 
As expected, 
the present results generally are in good agreement with the experimental observations \cite{lobovsky2014experimental} 
and previous numerical prediction \cite{rezavand2020weakly}. 
Compared with the experimental data \cite{lobovsky2014experimental}, 
a slightly faster wave front and a considerably higher backward wave are noted 
due to the inviscid model and the experimental uncertainty. 
Such discrepancies also have been observed in previous numerical studies in both single- and multi-phase simulations \cite{zhang2017weakly, rezavand2020weakly}. 
In addition, 
a convergence study of the multi-resolution simulation is also conducted herein.
In particular, 
the time evolution of water surface on a longer time scale
is evidently in good agreement with experiment with refined spatial resolutions. 
It is worth noting that the present prediction of water surface at $h_1$ shows a rapid convergence 
in capturing the correct temporal location, 
i.e., $\text t \left(\mathbf g/ \text H \right)^{0.5} = 5.5$, 
of the backward wave, 
which is not achieved in the previous studies to the best knowledge of the authors. 

\begin{figure}[htb!]
	\centering
	\includegraphics[trim = 2mm 2mm 2mm 2mm, clip, width=\textwidth]{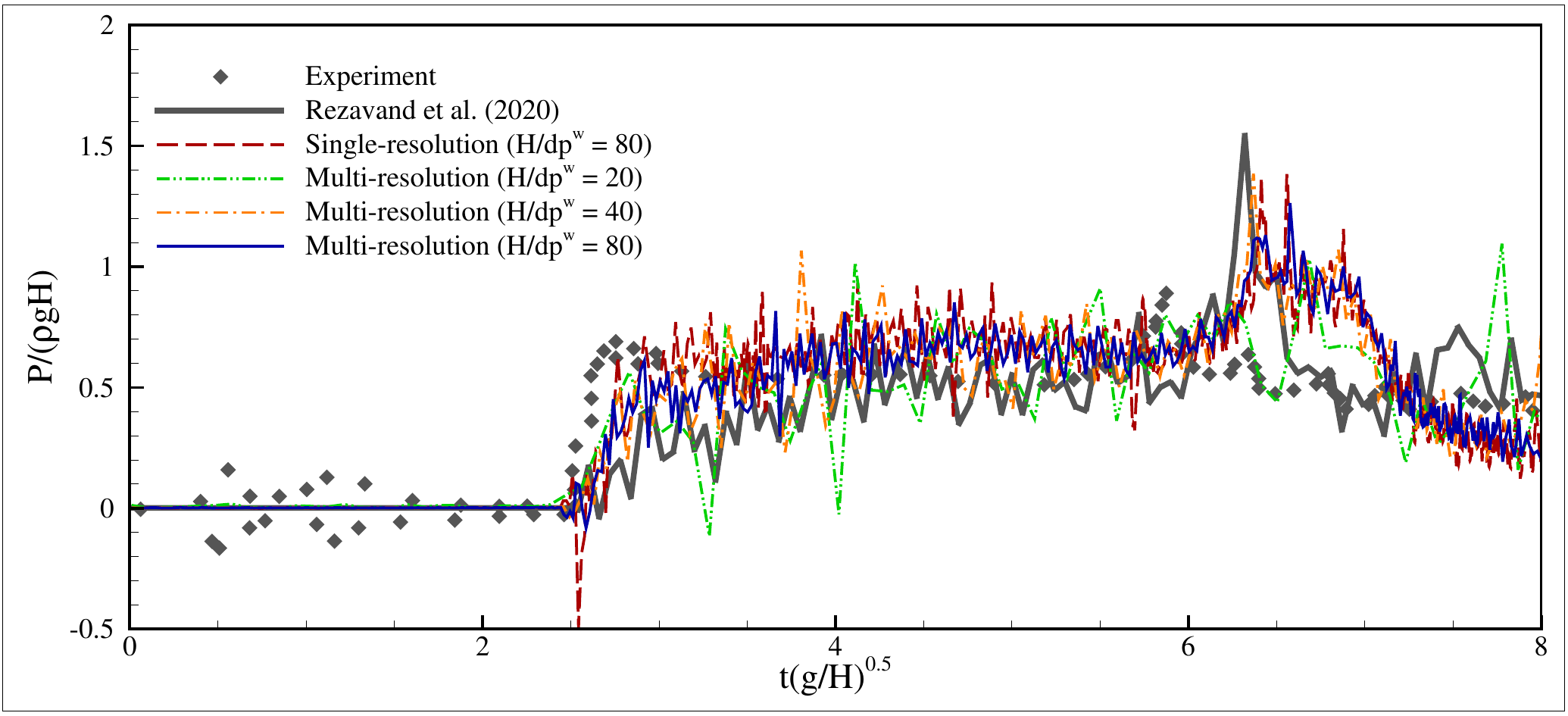}
	\caption{
		Dam-break flow: 
		The time history of the pressure signal probed by ensor $P$ in single- and multi-resolution simulations
		and its comparison with experimental data \cite{lobovsky2014experimental} 
		and numerical prediction \cite{rezavand2020weakly}. 
		Also,  
		a convergence study of the present multi-resolution method is conducted herein. 
		(For interpretation of the references to color in this figure legend, the reader is referred to the web version of this article.)
	}
	\label{figs:dam-pressure}
\end{figure}
Figure \ref{figs:dam-pressure} plots the time history of the pressure signals probed by sensor $P$ 
with the present method in single- and multi-resolution simulation 
and its comparison against experimental data \cite{lobovsky2014experimental} 
and numerical prediction obtained by the multi-phase SPH method \cite{rezavand2020weakly}. 
As expected, 
the main pressure plateau agrees well with the experimental \cite{lobovsky2014experimental} 
and numerical data \cite{lobovsky2014experimental}, 
while large oscillations mitigated with increased spatial resolution 
are exhibited due to the WC-assumption \cite{adami2012generalized, rezavand2020weakly, zhang2017weakly}. 

\begin{table}[htb!]
	\centering
	\caption{Dam-break flow: Computational efficiency. 
		To analyze the computational performance,  
		we evaluate the CPU wall-clock time of the computation of $H/dp^s = 80$ until $10$ dimensionless time instant. }
	\begin{tabular}{ccccc}
		\hline
		Cases      			& 	Single resolution & Multi resolution & Speedup \\
		\hline
		CPU time (s)	&   2856.5 & 461.5 & 6.19     \\ 
		\hline
	\end{tabular}
	\label{tab:dambreak-cputime}
\end{table}
Having the above qualitative and quantitative validations, 
we can conclude that the present multi-resolution method demonstrates 
its robustness and accuracy in capturing and maintaining sharp water-air interface 
without exhibiting unnatural voids and phase separation and predicting pressure signal 
during violent breaking and impact events,  
meanwhile achieves a computational speedup of $6.19$ compared with single-resolution counterpart as shown in Table \ref{tab:dambreak-cputime}.
\subsection{Nonlinear sloshing flow} \label{sec:sloshing}
The liquid sloshing phenomenon, 
which is characterized by violent and nonlinear surface motion induced by external excitation of the liquid container, 
occurs in various engineering applications, important examples include 
propellant sloshing in spacecraft tanks and rockets 
and liquid cargo sloshing in ships and trucks transporting (e.g. oil and liquefied natural gas). 
Extensive numerical studies have been conducted in literature \cite{dias2018slamming} 
to understand the pressure fluctuation near the surface,  
acoustic effects due to the strong impact 
and impacting loads on the container structure. 
These studies play a key role in preventing structure damage induced by sloshing flow, 
in particle when the tank motion frequency is close to the natural one of the inside liquid. 
Since its inception, 
the SPH method has been widely applied in the study of sloshing phenomenon with or without considering air phase
\cite{rafiee2011study, gotoh2014enhancement, shao2012improved, rafiee2012comparative}
due to its advantage of capturing violent breaking and impact events \cite{zhang2017weakly, zhang2022}. 
In this part, 
we consider two-dimensional two-phase sloshing flow to investigate the air effects on impacting pressure 
and assess the robustness and accuracy of the present method. 

\begin{figure}[htb!]
	\centering
	\includegraphics[trim = 0mm 1cm 0mm 2cm, clip,width=0.95\textwidth]{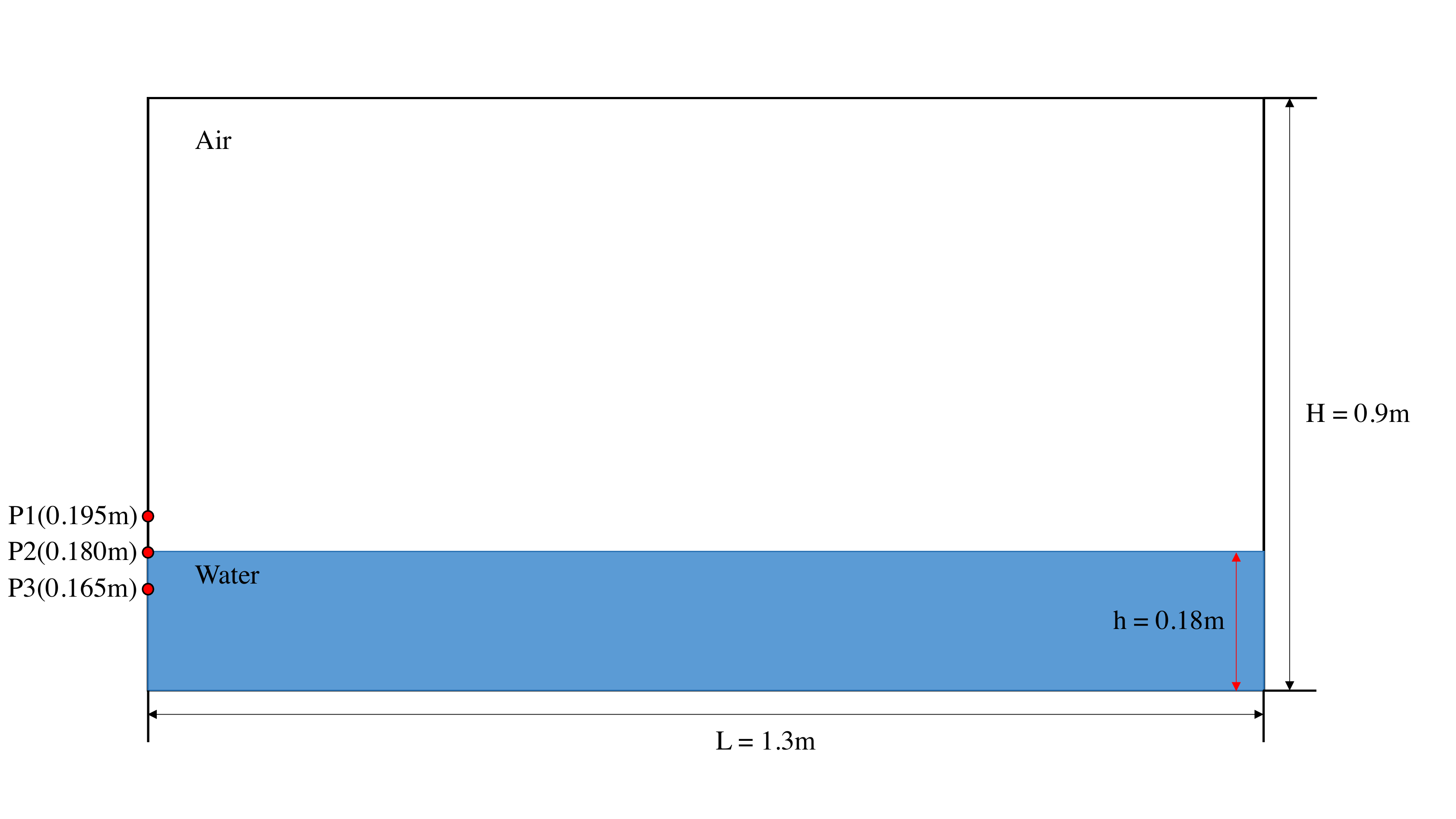}
	\caption{Nonlinear sloshing flow:
		Schematic illustration. 
		Note that there are three pressure sensors $P1$, $P2$ and $P3$ located on the left wall 
		to probe the pressure signals. 
	}
	\label{figs:sloshingsetup}
\end{figure}
Following the experimental study conducted by Rafiee et al. \cite{rafiee2011study, rafiee2012comparative}, 
where a rectangular tank with a low filling water is taken into consideration. 
The tank with length of $L=0.9~$m and height of $L=0.9~$m 
is partially filled with water of height $h=0.18~$m, 
while the remainder is filled by air, 
as in Figure \ref{figs:sloshingsetup}. 
The flow is considered to be inviscid herein and the density of water and air are set to 
$\rho_w = 1000~\mathrm{kg\cdot m^{-3}}$ and $\rho_a = 1~\mathrm{kg \cdot m^{-3}}$, respectively. 
The motion of the tank is driven by a sinusoidal excitation 
in $x$-direction of $x = A \sin(2.0 f \pi t)$, 
with $A = 0.1~m$ and $f = 0.496~s^{-1}$ respectively denoting the amplitude and frequency, 
implying a strong nonlinear violent surface motion  
as the excitation frequency $f$ is close to the natural one of the displaced water. 
For quantitative validation, 
three pressure sensors located on the left wall of the tank are applied to probe the impacting loads. 
To discretize the computational domain, 
the particles are placed on a regular lattice with a particle spacing of $dp = L/260$ 
and smoothing ratio $h^a = 2h^w$. 

\begin{figure}[htb!]
	\centering
	\includegraphics[trim = 1mm 1mm 1mm 1mm, clip,width=0.99\textwidth]{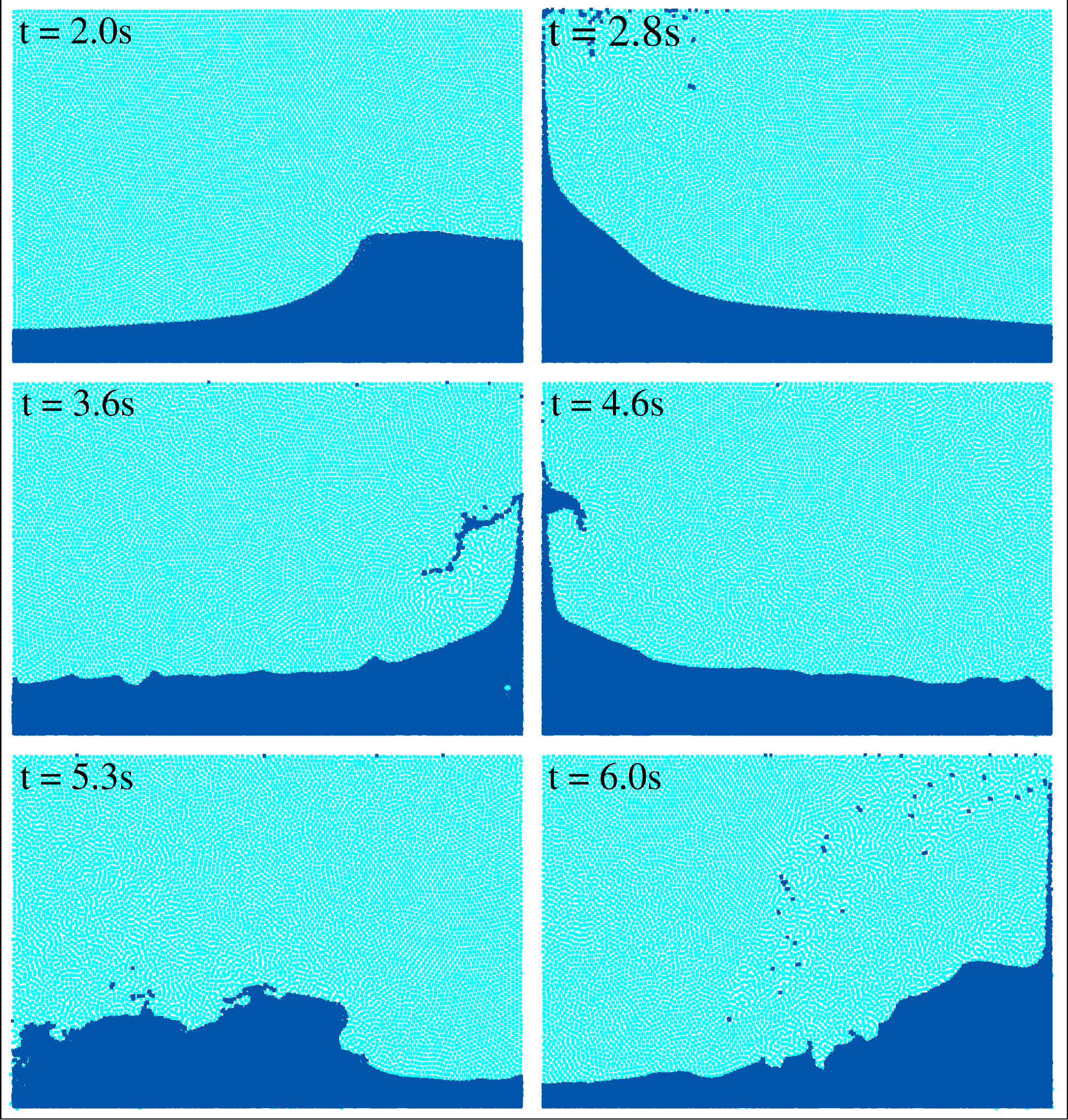}
	\caption{
		Nonlinear sloshing flow:
		Snapshots of particle distributions at different time instants.
		(For interpretation of the references to color in this figure legend, the reader is referred to the web version of this article.)	
	}
	\label{figs:sloshingsurface}
\end{figure}
Figure \ref{figs:sloshingsurface} portrays several snapshots of particle distributions with phase contour 
at different time instants in present multi-resolution simulation. 
Similar with the previous numerical results \cite{rezavand2020weakly, zhang2020dual, zhang2019weakly}, 
the main features of a nonlinear sloshing flow, 
i.e., a traveling wave with a crest, 
the resulting bore impact 
and breaking and the high run-up along the tank walls, 
are well captured by the present method.
With the presence of violent breaking, reentry and impact events of the water surface, 
the present multi-resolution simulation robustly captures and maintains a sharp water-air interface 
without exhibiting unnatural voids, unphysical phase separation and particle penetration, 
implying the performance of the proposed method in modeling complex and violent interface evolution. 

\begin{figure}[htb!]
	\centering
	\includegraphics[trim = 0mm 0mm 0mm 0mm, clip,width=\textwidth]{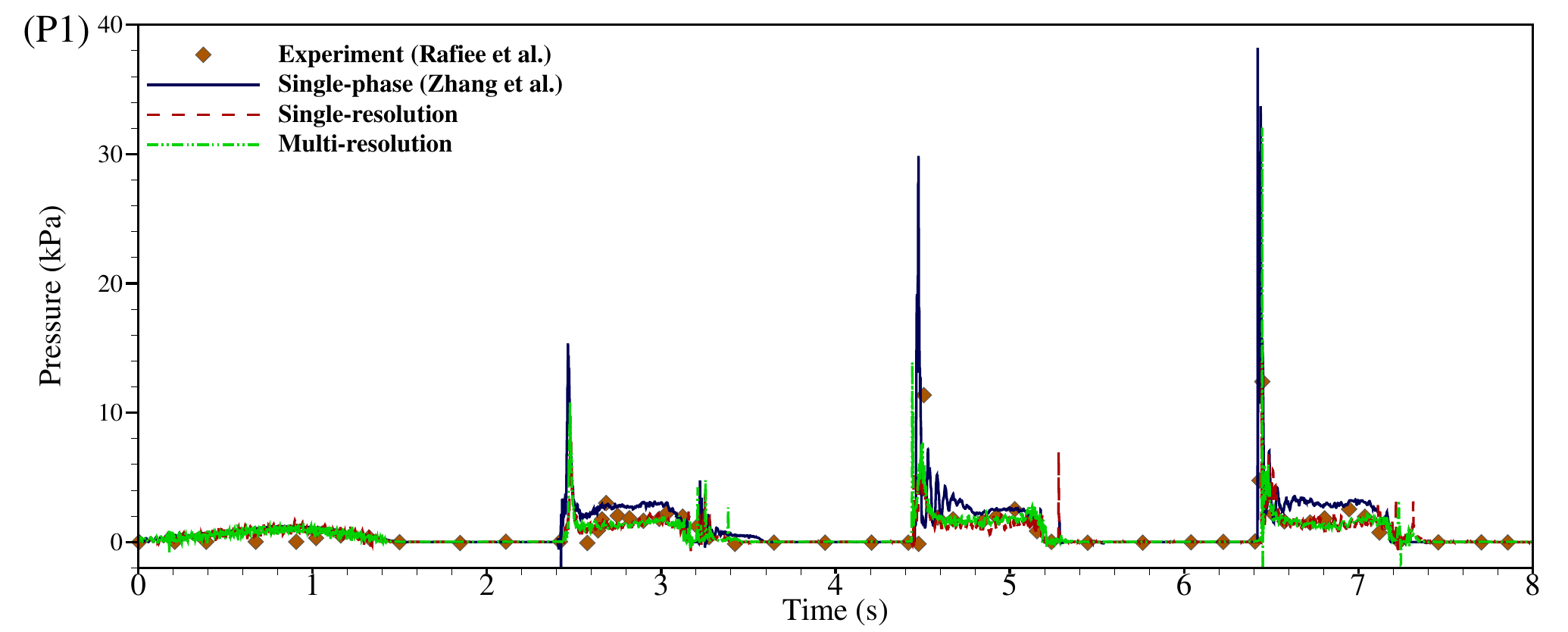}
	\includegraphics[trim = 0mm 0mm 0mm 0mm, clip,width=\textwidth]{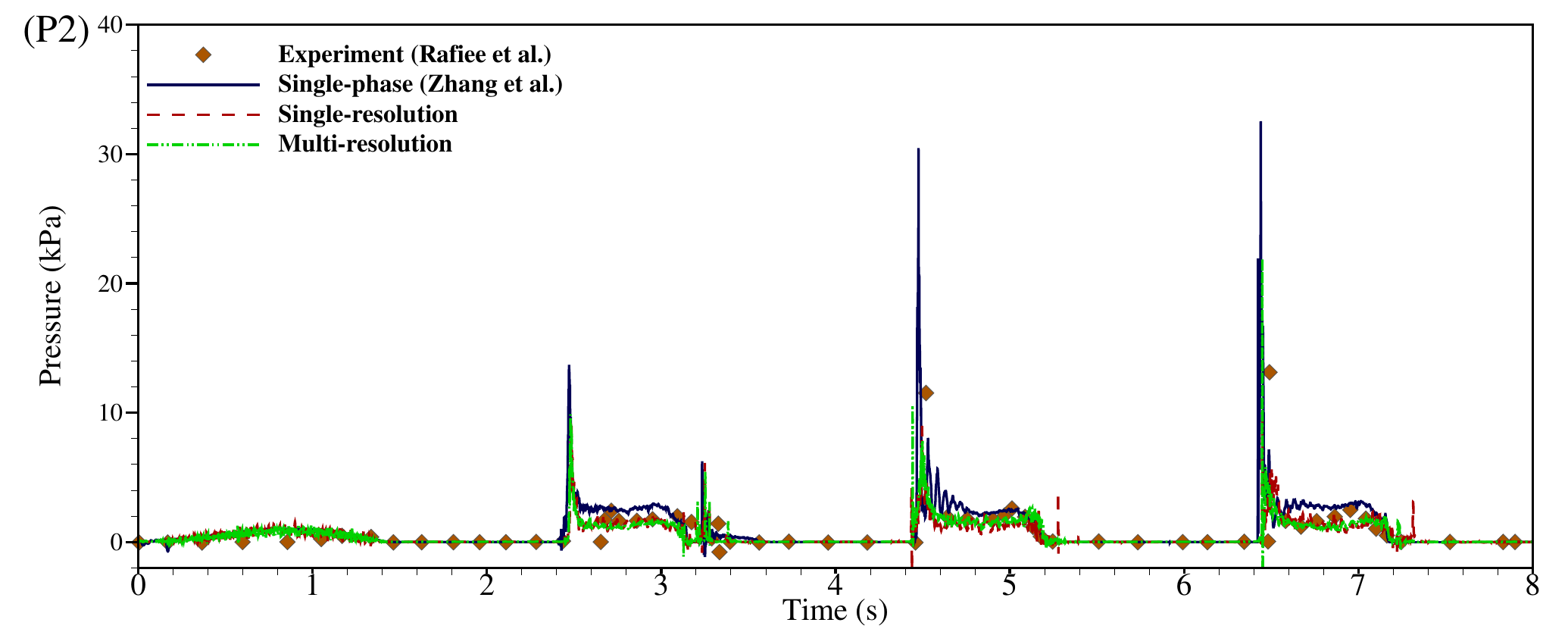}
	\includegraphics[trim = 0mm 0mm 0mm 0mm, clip,width=\textwidth]{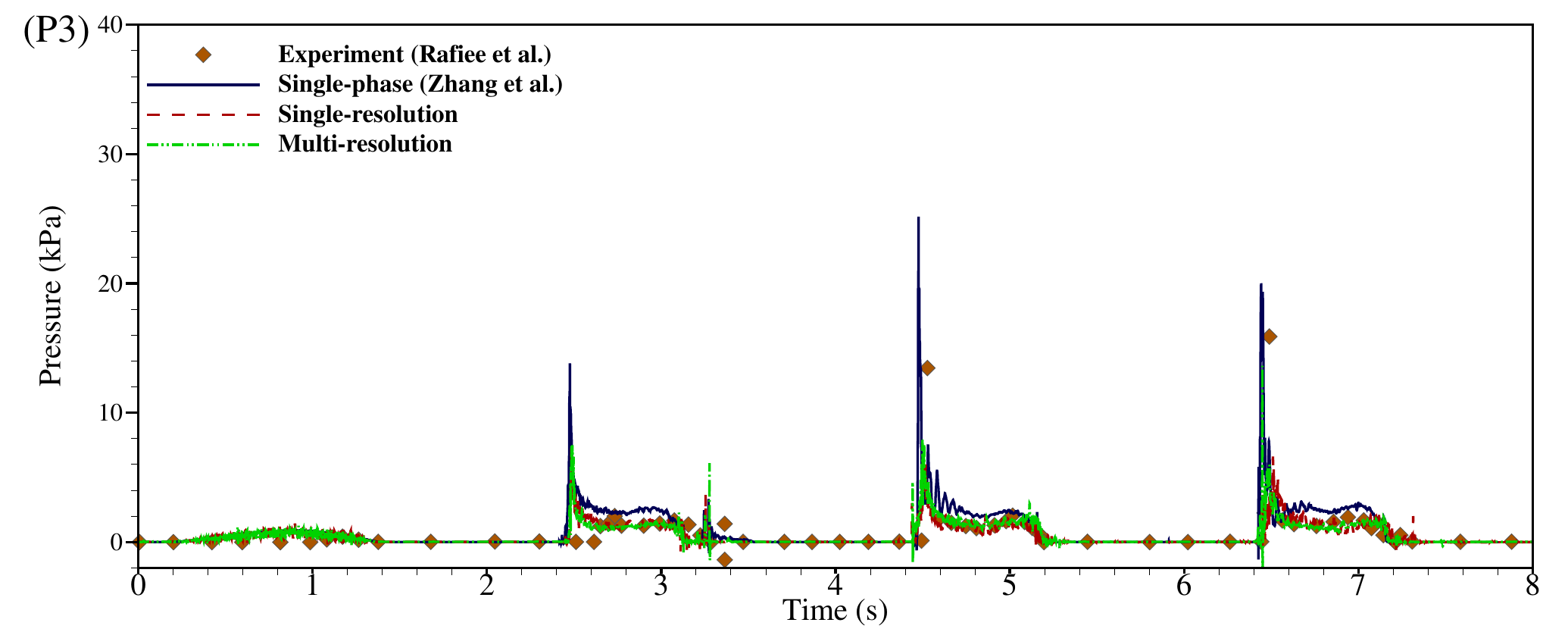}
	\caption{
		Nonlinear sloshing flow:
		The time history of the impact pressure signals probed by 
		sensors $P1$ $P2$ and $P3$, and its comparison against experimental data \cite{rafiee2011study} 
		and numerical results obtained by the single-phase SPH method \cite{zhang2020dual}. 
		(For interpretation of the references to color in this figure legend, the reader is referred to the web version of this article.)
	}
	\label{figs:sloshingpressure}
\end{figure}
To quantitatively assess the accuracy of the present method, 
Figure \ref{figs:sloshingpressure} depicts 
the time history of the numerical predicted impacting loads on the tank wall 
probed by sensors $P1$ $P2$ and $P3$ and its comparison against 
experimental data \cite{rafiee2011study} and 
numerical result obtained by the single-phase SPH method \cite{zhang2020dual}. 
Both present single- and multi-resolution results show a good agreement 
with the experimental signal despite of some oscillations induced by the WC-assumption. 
Note that there is noticeable difference between 
the impacting loads obtained by the simulations in single- and multi-resolution scenarios, 
indicating the accuracy of the present multi-resolution extension. 
Compared with the results obtained with single-phase SPH method \cite{zhang2020dual}, 
the present multi-phase simulations show improved accuracy in 
capturing the main pressure plateau with  
smaller peak pressure due to the existence of air cushion which moderates the impacting loads. 
Note that the compressibility of the air phase is neglected in the present simulation, 
while it complicates the evolution of the impacting loads which are strongly influenced by local phenomena, 
i.e., 
phase transition between liquid and vapor, 
liquid/gas mixture and surface tension, 
as noted by Dias and Ghidaglia \cite{dias2018slamming}. 

In this test, 
the present multi-resolution method demonstrates its 
robustness and accuracy in modeling nonlinear liquid sloshing phenomenon 
and the optimized computational efficiency by achieving a speedup of $23.76$ compared with the single-resolution counterpart as shown in Table \ref{tab:sloshing-cputime}.
\begin{table}[htb!]
	\centering
	\caption{Nonlinear sloshing flow: Analysis of computational efficiency. 
		The computer information is reported in Table \ref{tab:hydrostatic-fsi-cputime}. 
		For this test,  
		we evaluate the CPU wall-clock time for computation until $10.0~$s time instant.}
	\begin{tabular}{ccccc}
		\hline
		Cases      		& 	Single resolution & Multi resolution & Speedup \\
		\hline
		CPU time (s)	&   8708.5 & 366.5 & 23.76     \\ 
		\hline
	\end{tabular}
	\label{tab:sloshing-cputime}
\end{table}
%
\subsection{Dam-break flow through an elastic gate}\label{sec:mp-fsi}
\begin{figure}[htb!]
	\centering
	\includegraphics[trim = 1mm 3cm 1mm 3cm, clip,width=0.8\textwidth]{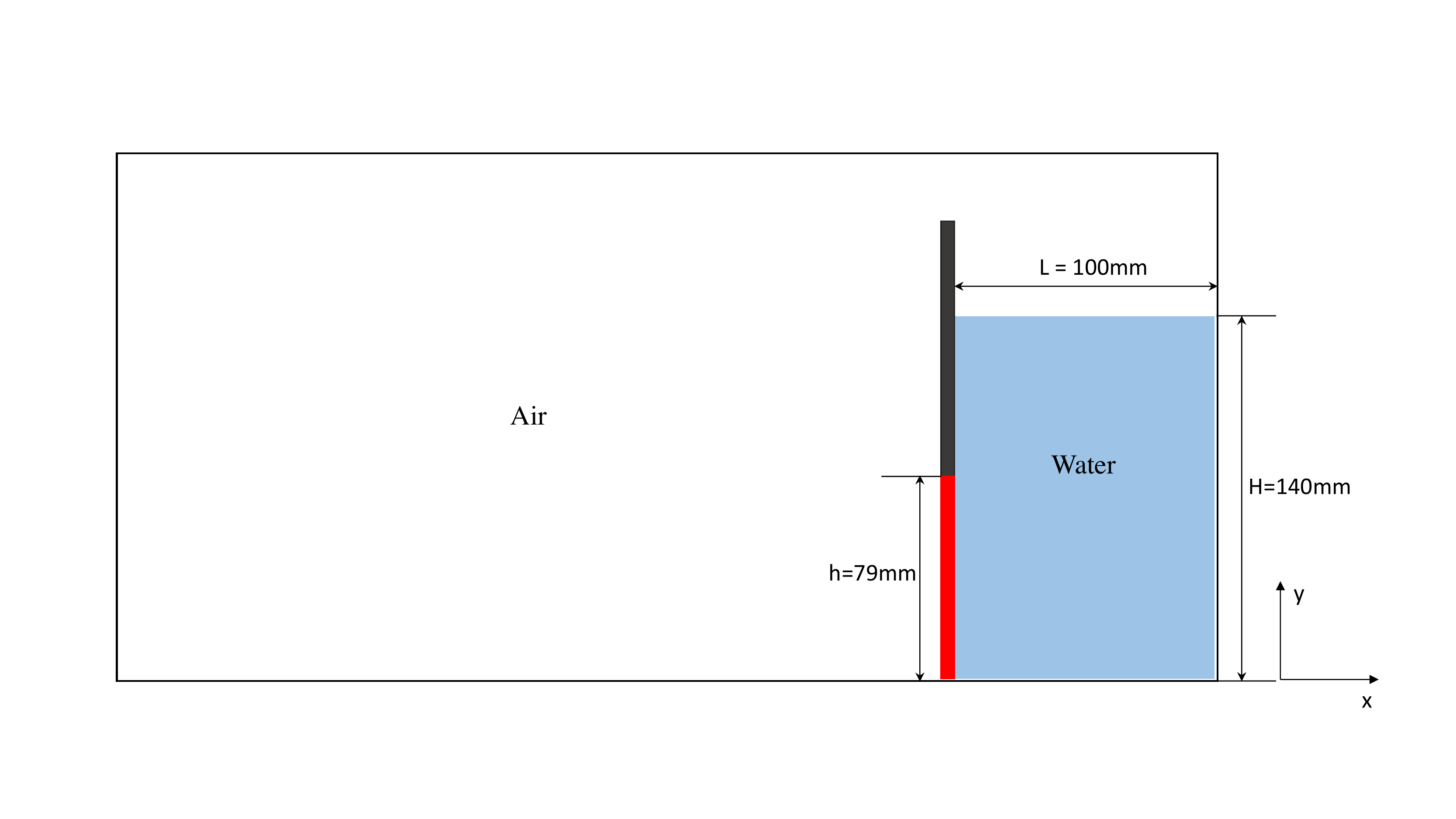}
	\caption{Dam-break flow through an elastic gate: 
		Schematic illustration. 
		Note that the width of the elastic gate is $b = 5mm$. }
	\label{figs:damplate-setup}
\end{figure}
In the following parts, 
we consider multi-phased hydroelastic FSI problem to validate the robustness, 
accuracy and efficiency of the present method in the simulation of multi-phase flow interacting with flexible structures. 
At first we consider the deformation of an elastic plate subjected to time-dependent hydrodynamics pressure 
induced by a two-phase dam-break flow with high density ratio. 
Following Ref. \cite{antoci2007numerical},
an elastic plate is displaced in a multi-phase flow induced by 
a bulk of water initially confined in a closed-air tank 
and clamped at its upper end and is free at the lower one as shown in Figure \ref{figs:damplate-setup}. 
A water column with height of $H = 0.14~\mathrm m$ and width of $L = 0.1~\mathrm m$
fills the space between the gate and right side wall of the tank whose remainder is filled with air.
Similar with Refs. \cite{antoci2007numerical, zhang2021multi}, 
the flow is considered to be inviscid with the water and air density $\rho_w = 1000~\mathrm{kg \cdot m^{-3}}$ 
and $\rho_a = 1~\mathrm{kg \cdot m^{-3}}$, resulting a large density ratio of $1000$. 
The material of the elastic gate is assumed to be linear isotropic material with density of $\rho_s = 1100~\mathrm{kg \cdot m^{-3}}$,
Young's modulus $E = 7.8~\mathrm{MPa}$ and Possion ratio $\nu = 0.47$. 
To impose the clam condition, 
the upper end of the gate is constrained by a rigid base.
For multi-resolution discretizations, 
the smoothing ratio is $h^a = 2h^w = 2h^s$.

\begin{figure}[htb!]
	\centering
	\includegraphics[trim = 1mm 1mm 1mm 1mm, clip,width=.95\textwidth]{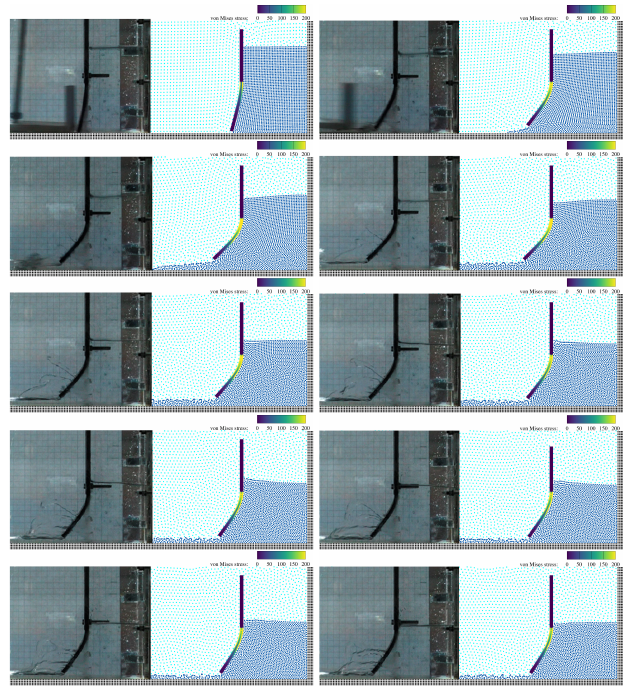}
	\caption{Dam-break flow through an elastic gate:
		Snapshots at every $0.04s$ starting from $t = 0.04s$ portray the deformed gate configurations 
		with von Mises stress contour and particle distributions with phase contour. 
		Here, the comparison is against experimental frames presented by Antoci et al. \cite{antoci2007numerical}. 
		(For interpretation of the references to color in this figure legend, the reader is referred to the web version of this article.)
	}
	\label{figs:dam-plate-surface}
\end{figure}
In Figure \ref{figs:dam-plate-surface}, 
we illustrate the gate deformation with von Mises stress contour 
and particle distribution with phase contour of the multi-resolution simulation 
at different time instants, and their comparison against the experimental observation of Antoci et al. \cite{antoci2007numerical}. 
Similar with Refs. \cite{antoci2007numerical, zhang2021multi},
these snapshots correspond to every $0.04~\mathrm{s}$ with the first frame at $t=0.04~\mathrm{s}$. 
Both the deformed configuration of the gate and the motion of the water surface 
are well predicted in the present simulation in comparison with the experimental frame. 
Also, 
the water-air interface is sharply captured and maintained during the whole dynamic process. 
In experiment, 
splashes are present due to the side leakage 
between the tank wall and the flexible plate \cite{antoci2007numerical, rafiee2009sph}, 
which is not exhibited in present two-dimensional simulation 
as those in Refs. \cite{rezavand2020weakly,zhang2021multi,zhang2022}

\begin{figure}[htb!]
	\centering
	\includegraphics[trim = 1mm 1mm 1mm 1mm, clip,width=.95\textwidth]{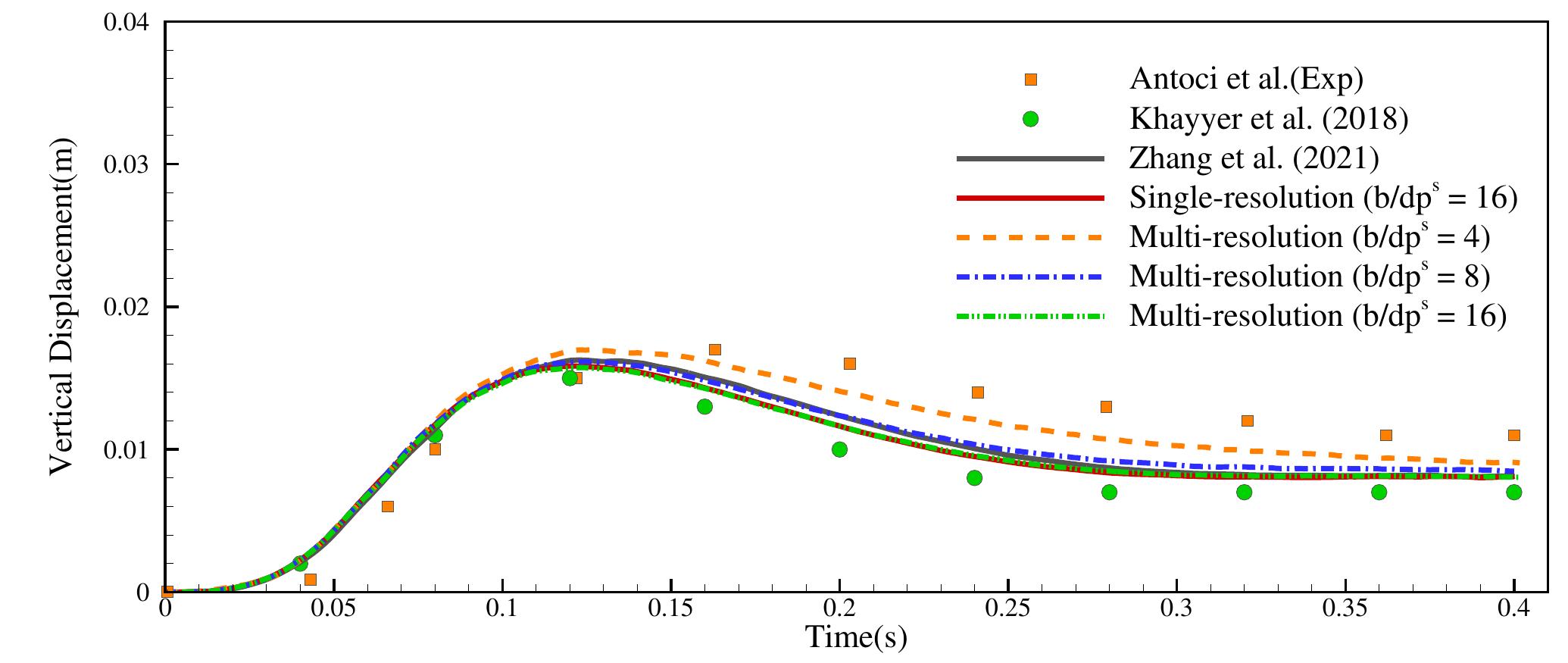}
	\includegraphics[trim = 1mm 1mm 1mm 1mm, clip,width=.95\textwidth]{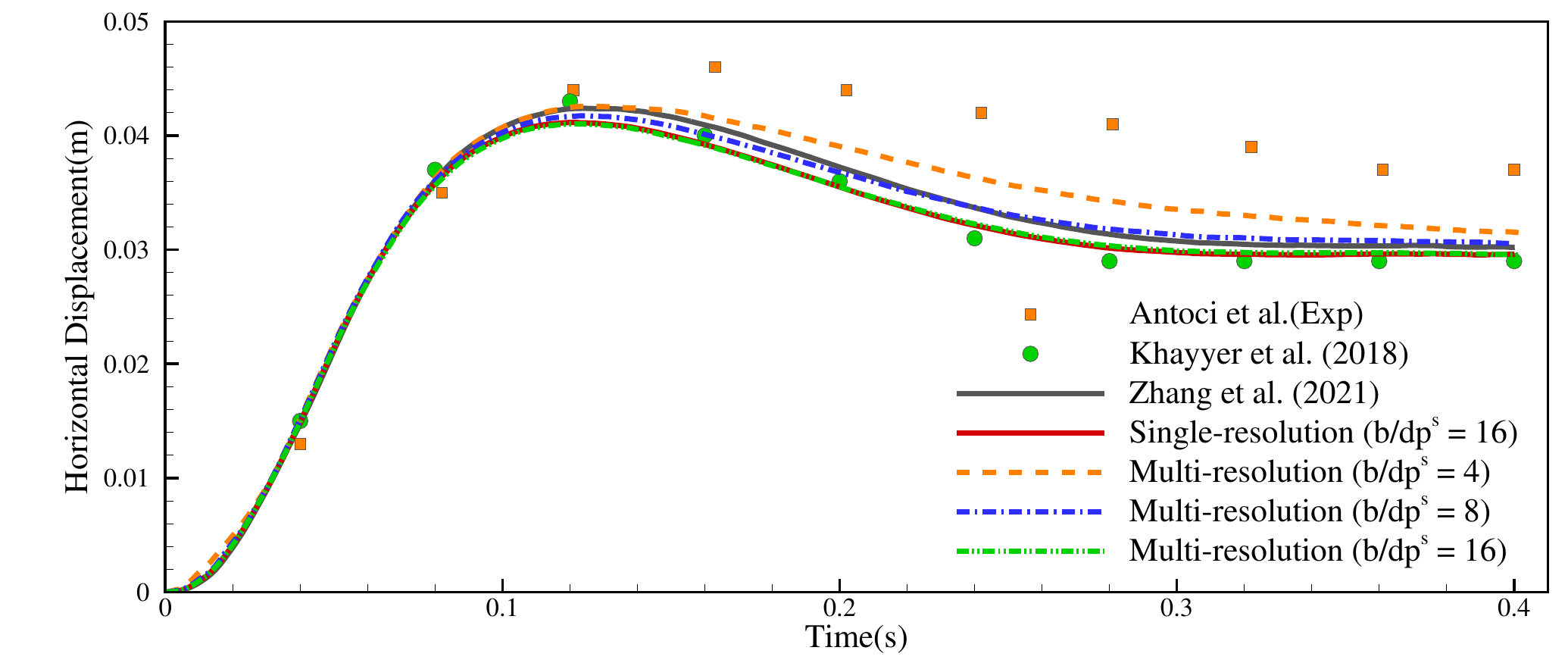}
	\caption{Dam-break flow through an elastic gate: 
		Vertical and horizontal displacements of the free gate end.
		Results obtained by both single- and multi-resolution scenarios are compared with experimental data reported by Antoci et al. \cite{antoci2007numerical},  
		numerical results using single-phase ISPH-SPH method of Khayyer et al. \cite{khayyer2018enhanced} (with Possion ratio $\nu= 0.4$)
		and single-phase SPH method of Zhang et al. \cite{zhang2021multi} (with Possion ratio $\nu= 0.47$). 
		For multi-resolution simulation, 
		a convergence study is also conducted herein by increasing the spatial resolution. 
		(For interpretation of the references to color in this figure legend, the reader is referred to the web version of this article.)
	}
	\label{figs:dam-plate-data}
\end{figure}
For quantitative validation, 
Figure \ref{figs:dam-plate-data} presents  
the horizontal and vertical displacements of the free gate end reproduced 
by the present method in both single- and multi-resolution scenarios and its comparison 
against the experimental data \cite{antoci2007numerical} 
and previous numerical results obtained by the ISPH-SPH method \cite{khayyer2018enhanced} and the SPH method \cite{zhang2021multi} 
without considering the air phase effects. 
In general, 
the present results show a good agreement with both experimental and numerical data. 
More specifically, 
the gate deformation at the early stage ($t < 0.1~\mathrm s$) is well predicted by numerical models, 
while is underestimated afterward  compared with experimental observations. 
This discrepancy is associated with the applied material model as noted by Yang et al. \cite{yang2012free}. 
Figure \ref{figs:dam-plate-data} also provides a convergence study of the present multi-resolution method .
As the increase of the spatial resolution, 
the deformation difference between two refinement level is rapidly decreased. 
Note that the present converged value of the gate displacement is very close to those presented in literature \cite{khayyer2018enhanced, zhang2021multi}

Having the qualitative and quantitative validations in hand, 
we can conclude that the present multi-resolution method can accurately predict the hydroelastic response of 
a elastic structured displaced in a multi-phase flow, 
as well as achieves a computational speedup of $4.45$ 
compared with the single-resolution counterpart as shown in Table \ref{tab:damgate-cputime}.
\begin{table}[htb!]
	\centering
	\caption{Dam-break flow through an elastic gate: Analysis of computational efficiency. 
		The computer information is reported in Table \ref{tab:hydrostatic-fsi-cputime}. 
		For this test,  
		we evaluate the CPU wall-clock time for computation until $0.04~$s time instant.}
	\begin{tabular}{ccccc}
		\hline
		Cases      		& 	Single resolution & Multi resolution & Speedup \\
		\hline
		CPU time (s)	&   3012.12 & 676.72 & 4.45     \\ 
		\hline
	\end{tabular}
	\label{tab:damgate-cputime}
\end{table}
%
\subsection{Dam-break flow impacting an elastic plate}\label{sec:dambreak-fsi} 
\begin{figure}[htb!]
	\centering
	\includegraphics[trim = 1cm 4cm 1cm 4cm, clip, width= \textwidth]{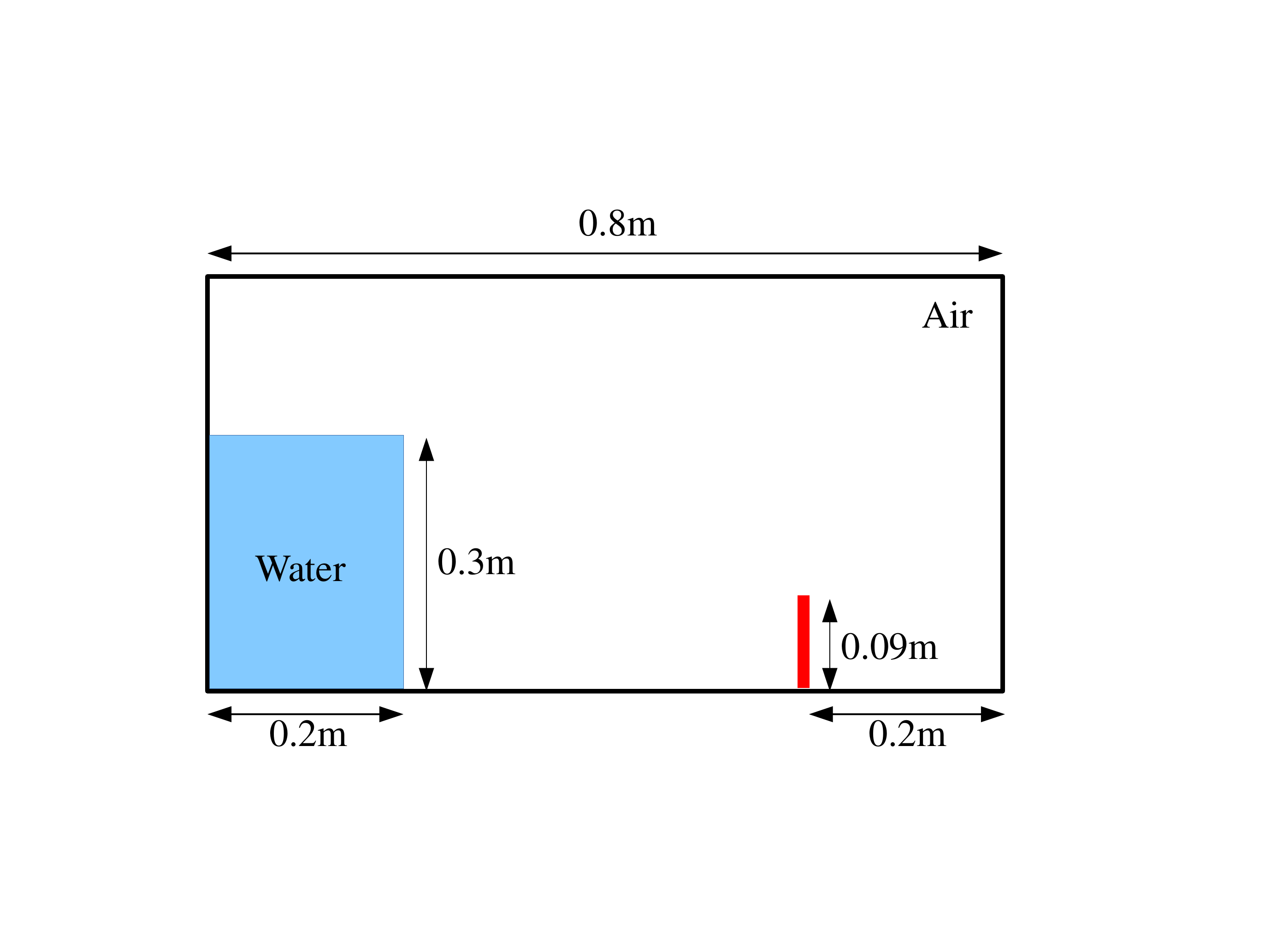}
	\caption{
		Dam-break flow impacting an elastic plate: Schematic illustration. 
		Note that the thickness of the flexible structure is $b = 0.004~m$ 
		and three markers with locations $h = 0.0875 ~m$ (Marker $1$) from the tan bottom, 
		$0.065 ~m$ (Marker $2$) and $0.04 ~m$  (Marker $3$) are set for probing the horizontal displacement of the flexible plate.}
	\label{figs:mp-dam-fsi-setup}
\end{figure}
In previous parts, 
the present method demonstrates its robustness and accuracy for modeling multi-phase flow with 
high-density ratio and violent interface evolution events such as breaking and impact 
and hydroelstics multi-phased FSI problems. 
To further demonstrate its performance for violent multi-phase flow interacting with flexible structure, 
we consider herein a two-dimensional two-phase dam-break flow impacting on an elastic plate.
This example has been studied as a benchmark test with both experimental and numerical data \cite{liao2015free, sun2019study, zhang2022} 
available for qualitative and quantitative validations.  
Following the experimental setup in Ref. \cite{liao2015free}, 
the dam-break flow is activated by 
a column of water with the size of height $H_w = 0.3~\mathrm{m}$ and length $L_w = 0.2~\mathrm{m}$ initially located at 
the left side of a tank with length $L_t = 0.8~\mathrm{m}$ and height $H_t = 0.6~\mathrm{m}$ whose reminder is filled with air, 
as portrayed in Figure \ref{figs:mp-dam-fsi-setup}. 
At the flow downstream, 
an elastic plate with the size of height $h = 0.09~\mathrm{m}$ and thickness $b = 0.004~\mathrm{m}$ 
is displaced $0.2~\mathrm{m}$ far away from the right side wall of tank, 
as shown in Figure \ref{figs:mp-dam-fsi-setup}. 
Similar with Refs. \cite{liao2015free, sun2019study, zhang2022},
the flow is considered to be inviscid and 
the density of water and air is respectively set as $\rho_w = 997~\mathrm{kg \cdot m^{-3}}$ 
and $\rho_a = 1.225~\mathrm{kg \cdot m^{-3}}$, 
resulting a relatively high density ratio about $1000$. 
As for the elastic plate, 
we consider a rubber-like material with density $\rho_s = 1161.5~\mathrm{kg \cdot m^{-3}}$,
Young's modulus $E = 3.5~\mathrm{MPa}$ and Poisson ratio $\nu = 0.45$. 
To quantitatively validate the accuracy, 
the horizontal displacement of the plate is measured by 
three markers on its middle line with different initial heights,  
i.e., 
Marker $1$ at $h_1 = 0.0875 ~\mathrm m$ from the tank bottom, 
Marker $2$ at $h_1 = 0.065 ~\mathrm m$ and Marker $3$ at $h_1 = 0.04 ~\mathrm m$, 
as Refs. \cite{liao2015free, sun2019study, zhang2022}. 
In this paper, 
we consider two cases with different smoothing ratios, 
viz., Case-I of $dp^a = 2.0 dp^w = 2.0 dp^w$ and Case-II of $dp^a = dp^w = dp^w$, 
and both with the initial particle spacing of $dp^s = b / 4$. 

\begin{figure}[htb!]
	\centering
	\includegraphics[trim = 1mm 1mm 1mm 1mm, clip, width=0.485\textwidth]{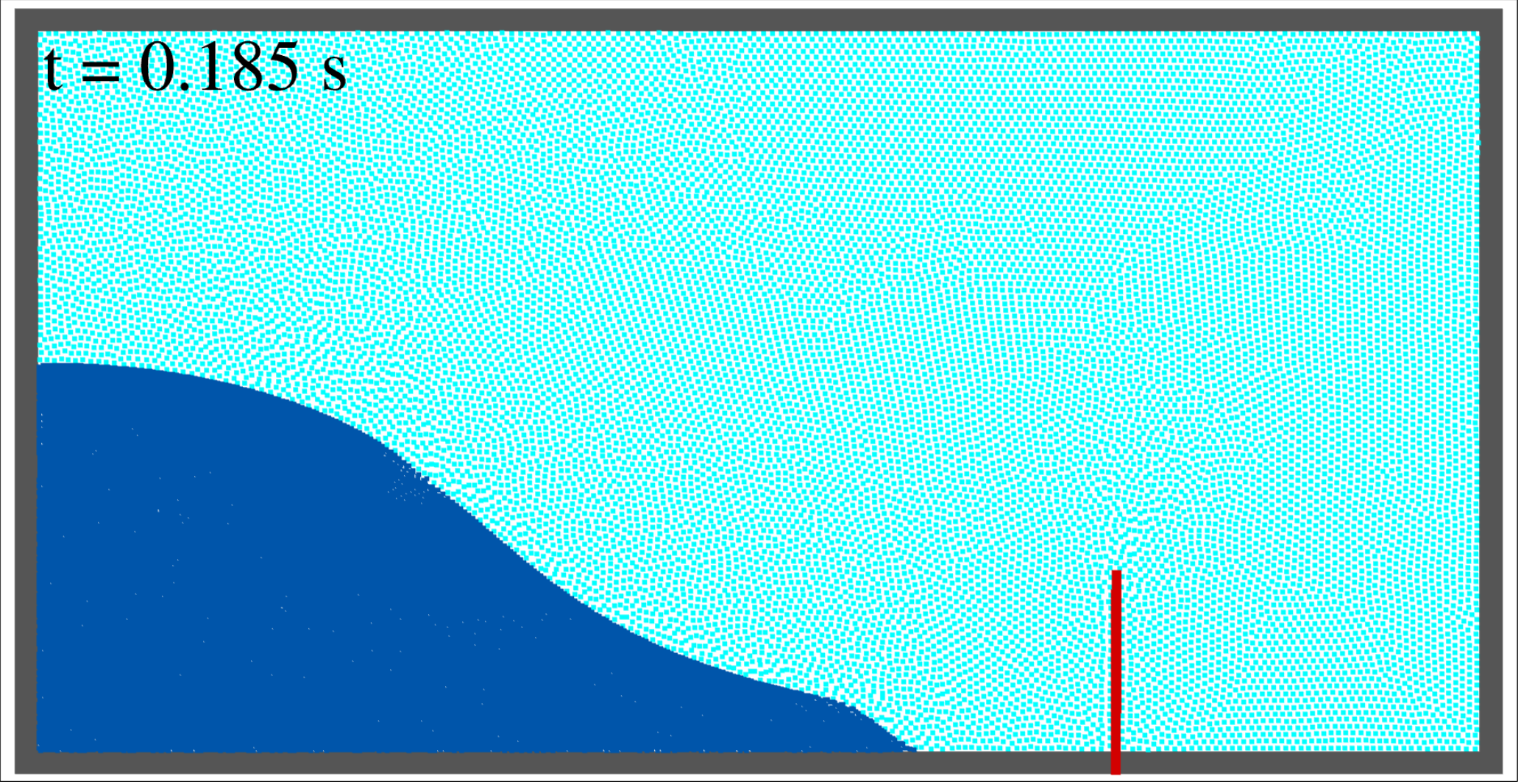}
	\includegraphics[trim = 1mm 1mm 1mm 1mm, clip, width=0.485\textwidth]{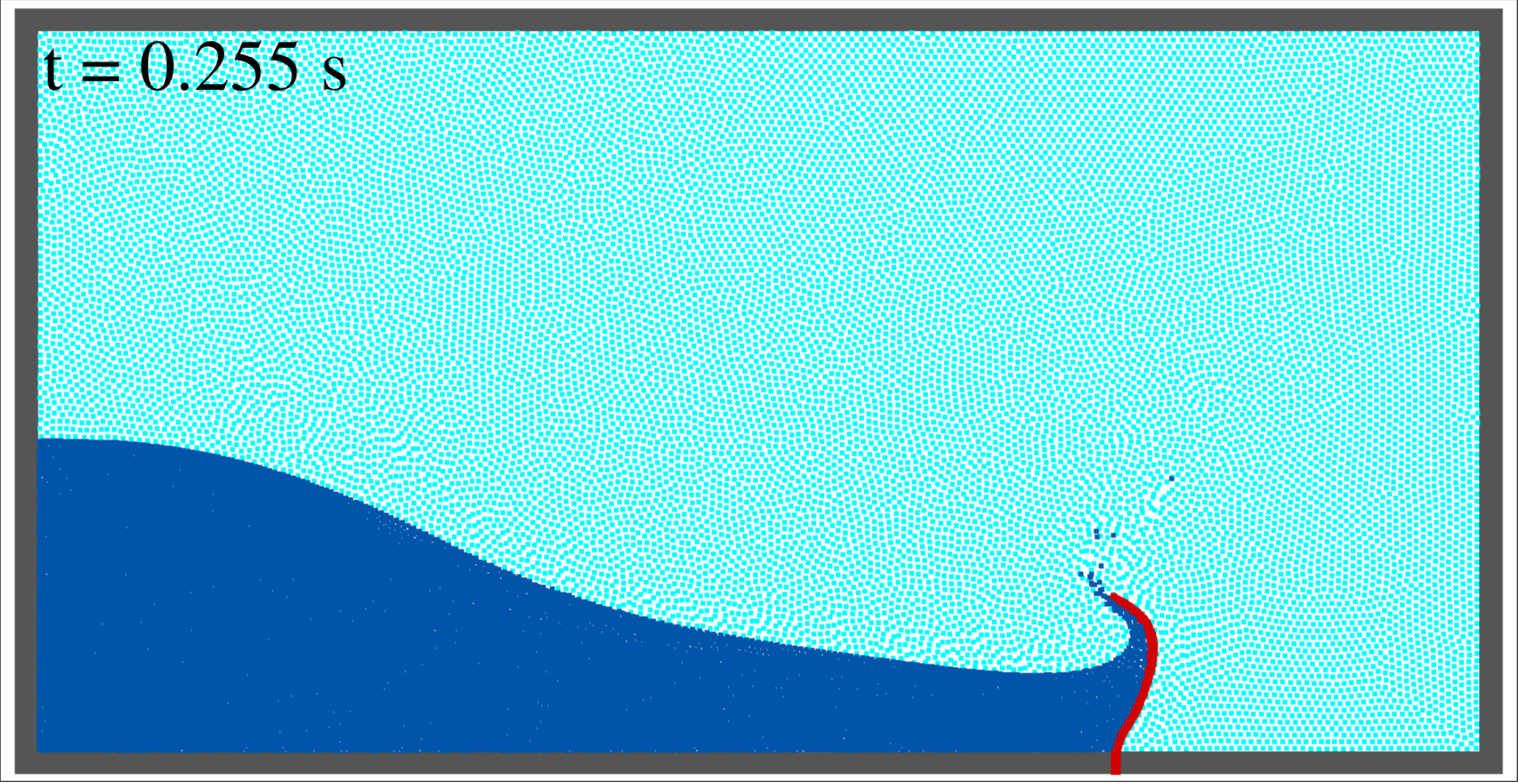}\\
	\includegraphics[trim = 1mm 1mm 1mm 1mm, clip, width=0.485\textwidth]{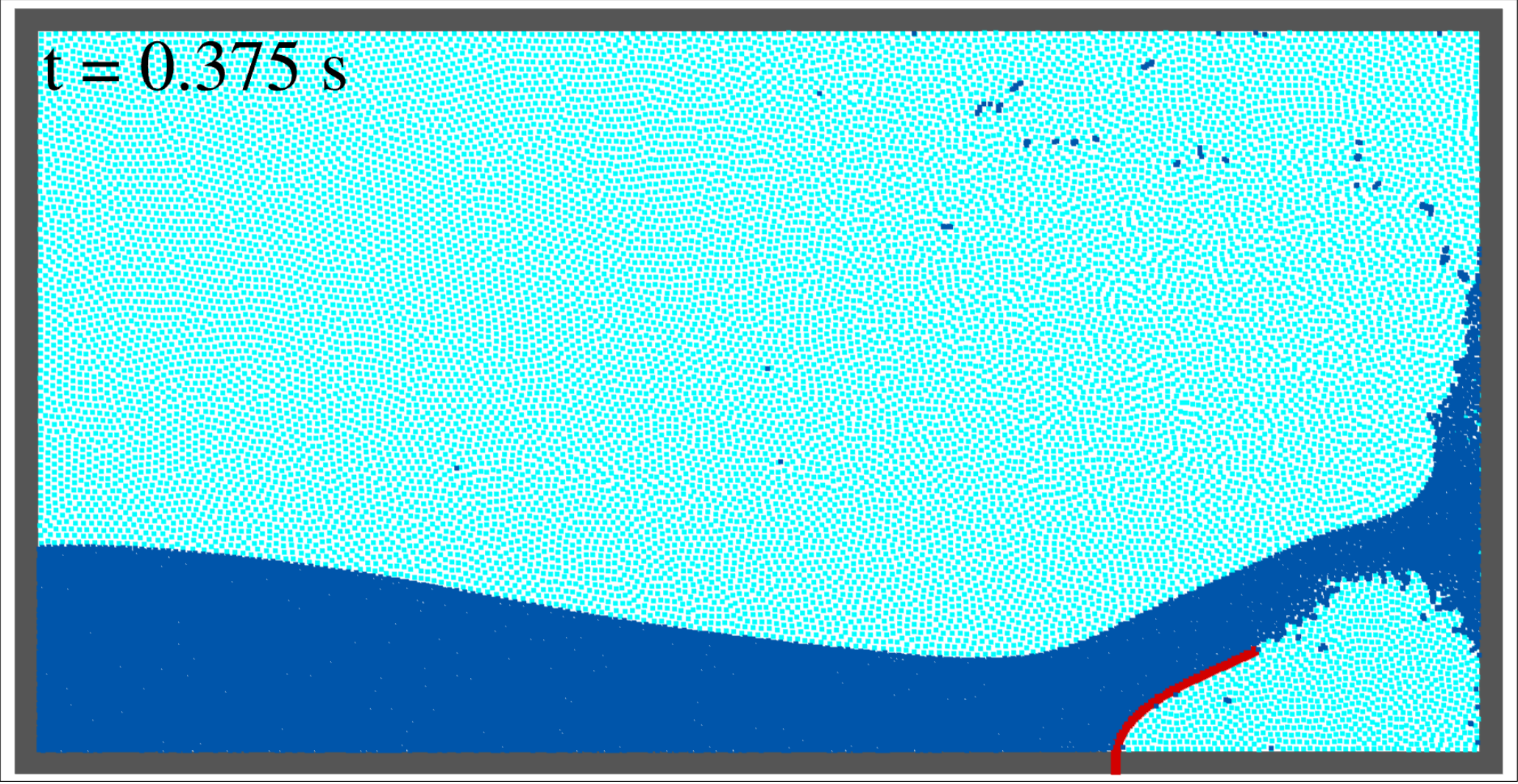}
	\includegraphics[trim = 1mm 1mm 1mm 1mm, clip, width=0.485\textwidth]{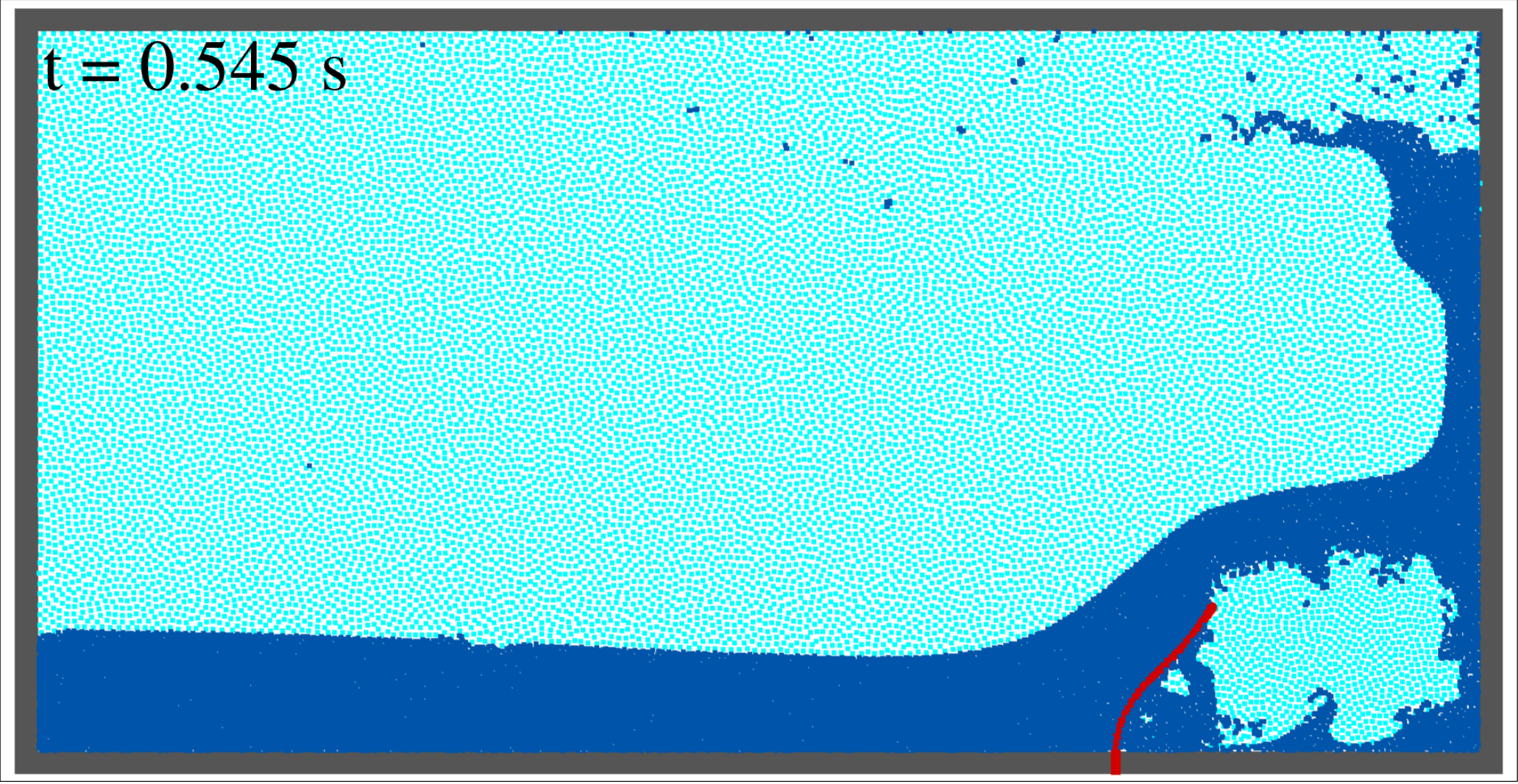}
	\caption{
		Dam-break flow impacting an elastic plate: 
		Snapshots of the water-air interface evolution 
		and the plate deformation at different time instants for Case-I.
	}
	\label{figs:mp-dam-fsi}
\end{figure}
Figure \ref{figs:mp-dam-fsi} depicts snapshots of the water-air interface evolution 
and the plate deflection at typical time instants for Case-I where the discretization is conducted in multi-resolution scenario. 
Before the water front impacting on the plate, 
a typical dam-break flow is observed as the flow propagates in the tank. 
At the impacting stage, 
the flow firstly bends the elastic plate, 
then overflows it and finally impacts on the right side wall of the tank, 
inducing two strong impact phenomena on both rigid and flexible structures. 
Similar with the previous numerical results \cite{sun2019study, zhang2022}, 
the featured phenomena are well captured in the present multi-resolution simulation. 
Also, 
the water-air interface is well captured and sharply maintained without exhibiting air particle penetration 
at both impact phenomena, 
demonstrating the robustness of present method. 
Note that as the results of Case-II do not exhibit visible difference from those reported in Figure \ref{figs:mp-dam-fsi} 
we refrain from showing them additionally. 

\begin{figure}[htb!]
	\centering
	\includegraphics[trim = 1mm 1mm 1mm 1mm, clip, width=0.99\textwidth]{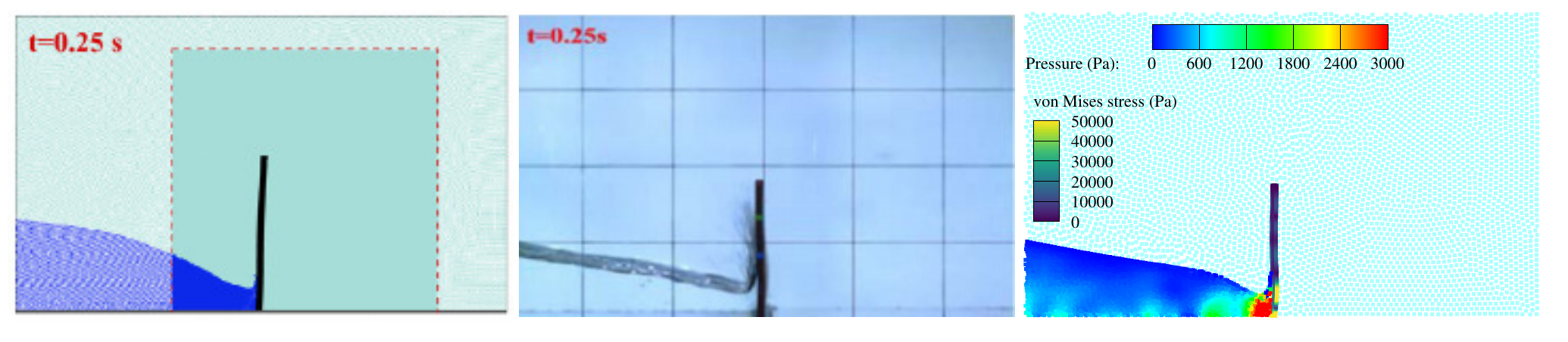}
	\includegraphics[trim = 1mm 1mm 1mm 1mm, clip, width=0.98\textwidth]{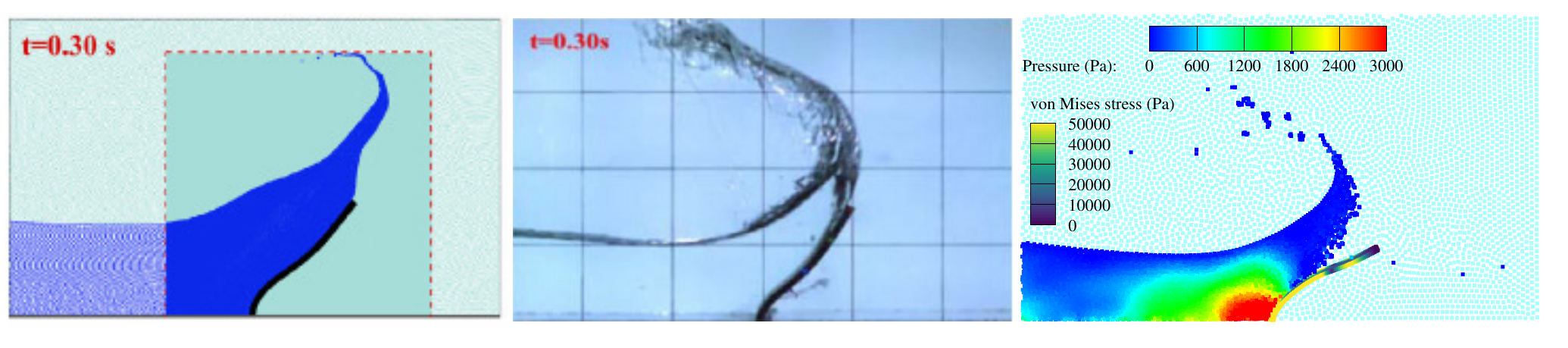}
	\includegraphics[trim = 1mm 1mm 1mm 1mm, clip, width=0.98\textwidth]{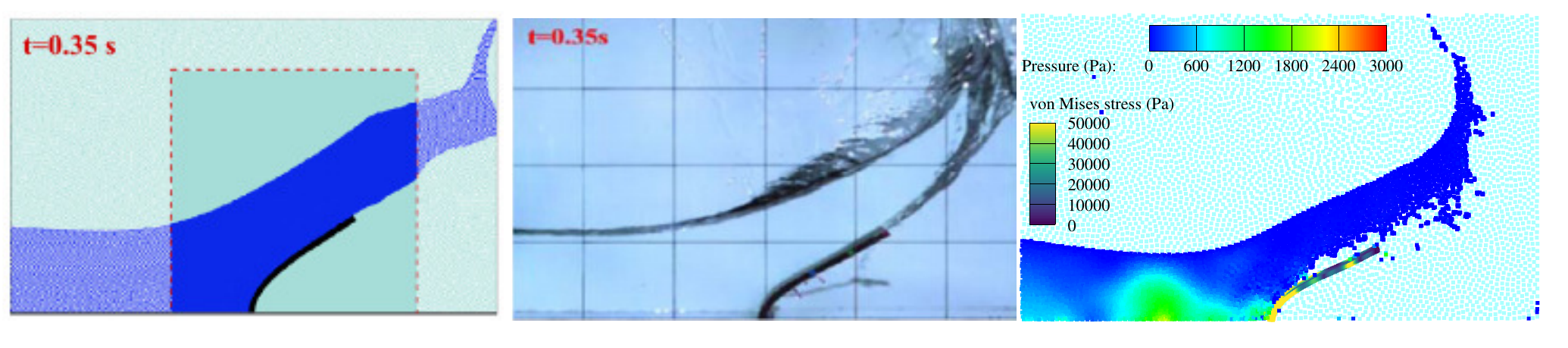}
	\includegraphics[trim = 1mm 1mm 1mm 1mm, clip, width=0.99\textwidth]{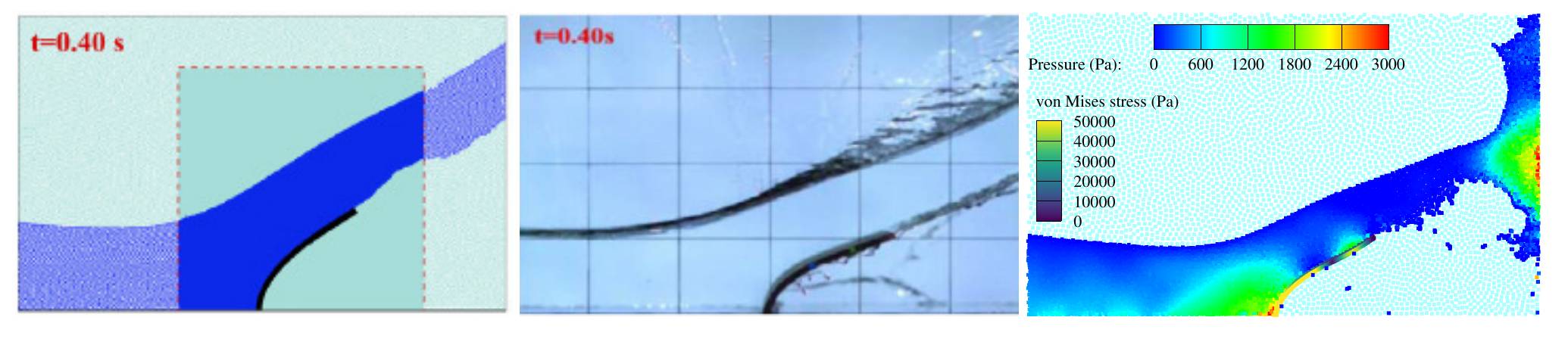}
	\includegraphics[trim = 1mm 1mm 1mm 1mm, clip, width=0.99\textwidth]{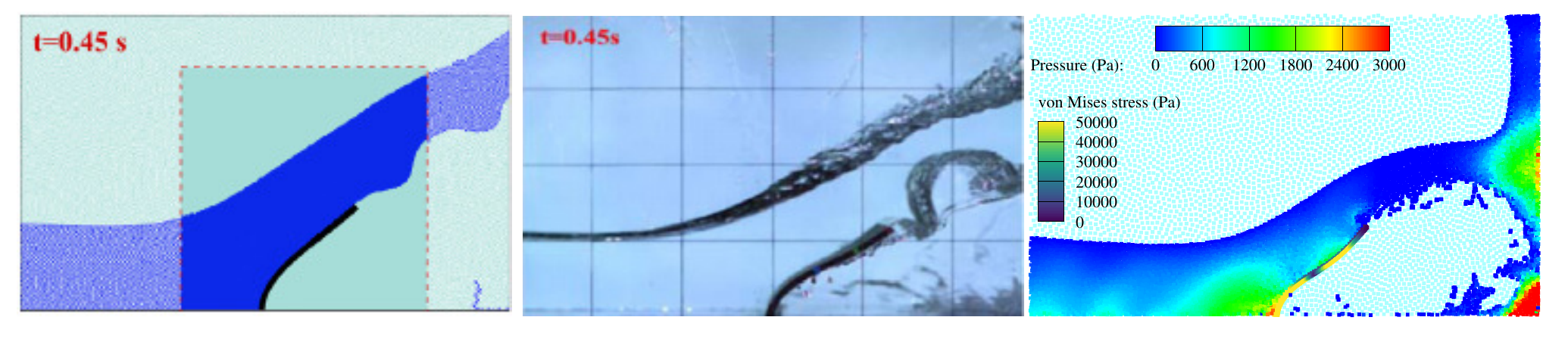}
	\caption{
		Dam-break flow impacting an elastic plate: 
		Snapshots at different time instants of particle distribution with pressure contour  
		and the deformed configuration with von Mises stress contour of the plate at different time instants for Case-I. 
		Here, the qualitative comparison is against experimental frames of Liao \etal \cite{liao2015free} 
		and numerical results obtained by multi-phase $\delta$-SPH method of Sun \etal \cite{sun2019study}.  
		(For interpretation of the references to color in this figure legend, the reader is referred to the web version of this article.)
	}
	\label{figs:mp-dam-fsi-exp}
\end{figure}
Figure \ref{figs:mp-dam-fsi-exp} portrays the particle distribution with pressure contour  
and the deformed configuration with von Mises stress contour of the plate at different time instants for Case-I, 
and the comparison against experimental observations presented by Liao \etal \cite{liao2015free} 
and numerical results obtained by multi-phase $\delta$-SPH method by Sun \etal \cite{sun2019study}. 
In general, 
a qualitative good agreement on the interface evolution and the plate deflection is noted. 
Compared with the experimental observations \cite{liao2015free},  
the water-air interface evolution and the plate deflection are reasonably predicted by the present multi-resolution method 
similar as the FEM-DSM \cite{liao2015free} 
and the multi-phase SPH methods \cite{sun2019study, zhang2022}.  
The main features observed in the experiment, 
i.e., a large splash jet induced by the water front impact and overflow, 
the disturbance of the cavity boundary induced by the plate vibration 
and the water droplets in the right corner after the secondary impact, 
are well captured. 
However, 
some discrepancies between the numerical results and the experimental observation are noted.
For example the numerical predicted overflow front composed by the splashing droplets is slightly smaller 
than the experimental observation due to the dominant three-dimensional effect 
and the overflow jet shows mistier water-air-interface after overflowing the plate 
due to the induced vibration, which will be shown in the following part of quantitative validation. 
These discrepancies are also observed in the numerical results obtained by different methods 
in literature \cite{sun2019study, liao2015free, zhang2022}. 
Compared with that obtained by multi-phase $\delta-$SPH method of Sun \etal \cite{sun2019study}, 
the present one exhibits strong disturbance in the cavity boundary of the water overflow jet  
and noticeable separation between the jet and plate at time instants $t = 0.3~\mathrm s$ and $t = 0.35~\mathrm s$ 
induced by the model difference. 
In Ref. \cite{sun2019study}, 
the multi-phase $\delta$-SPH is applied 
with assigning the air an artificial speed of sound $10$ times larger than the one of the water,
adding backgroud pressure in the EoS of both phases 
and implementing multi-phase tensile instability control. 
The present method \cite{rezavand2020weakly} 
applies a more simple and efficient model, 
where same value is assigned to the artificial speed of sound for both phases and 
the backgroud pressure is only applied to the transport-velocity of air.

\begin{figure}[htb!]
	\centering
	\includegraphics[trim = 1mm 1mm 1mm 1mm, clip, width=0.85\textwidth]{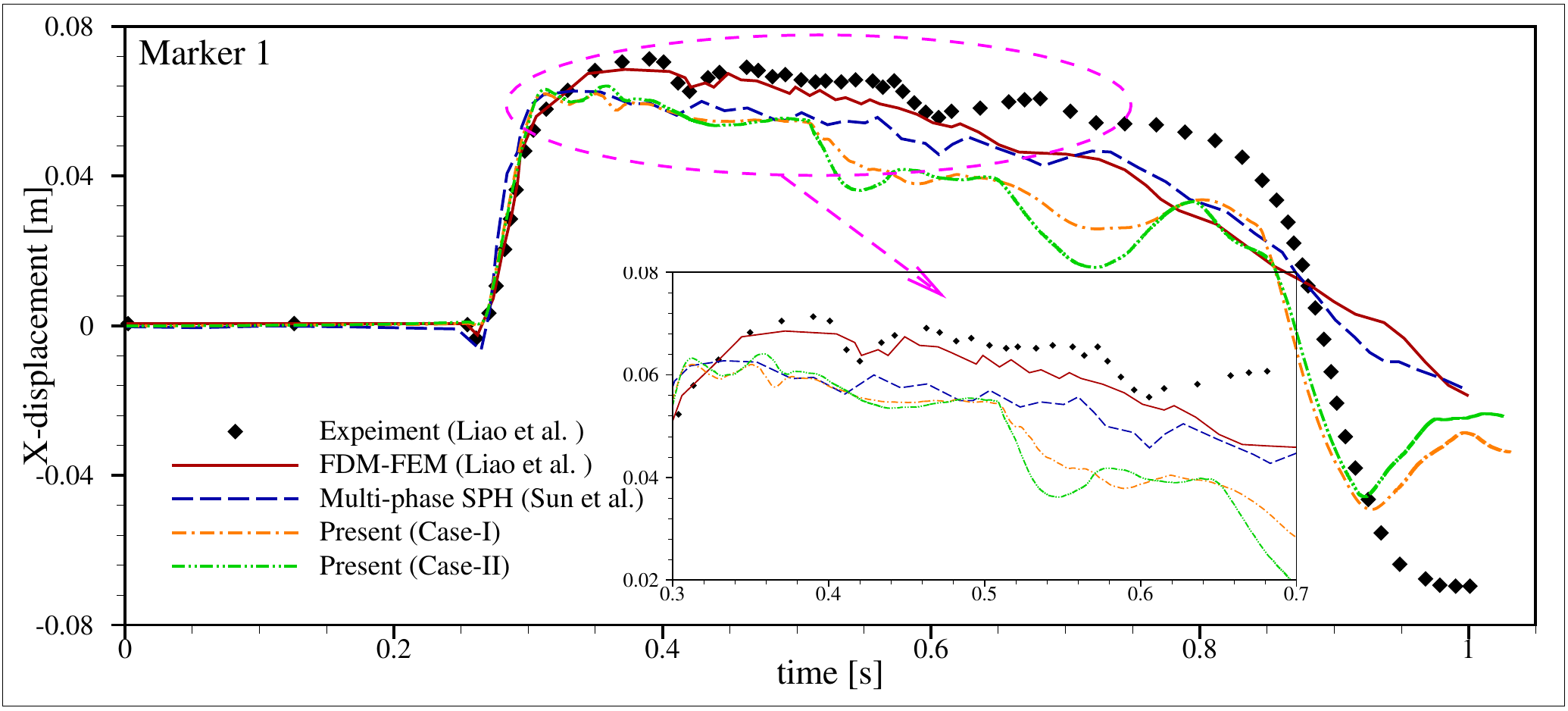}
	\includegraphics[trim = 1mm 1mm 1mm 1mm, clip, width=0.85\textwidth]{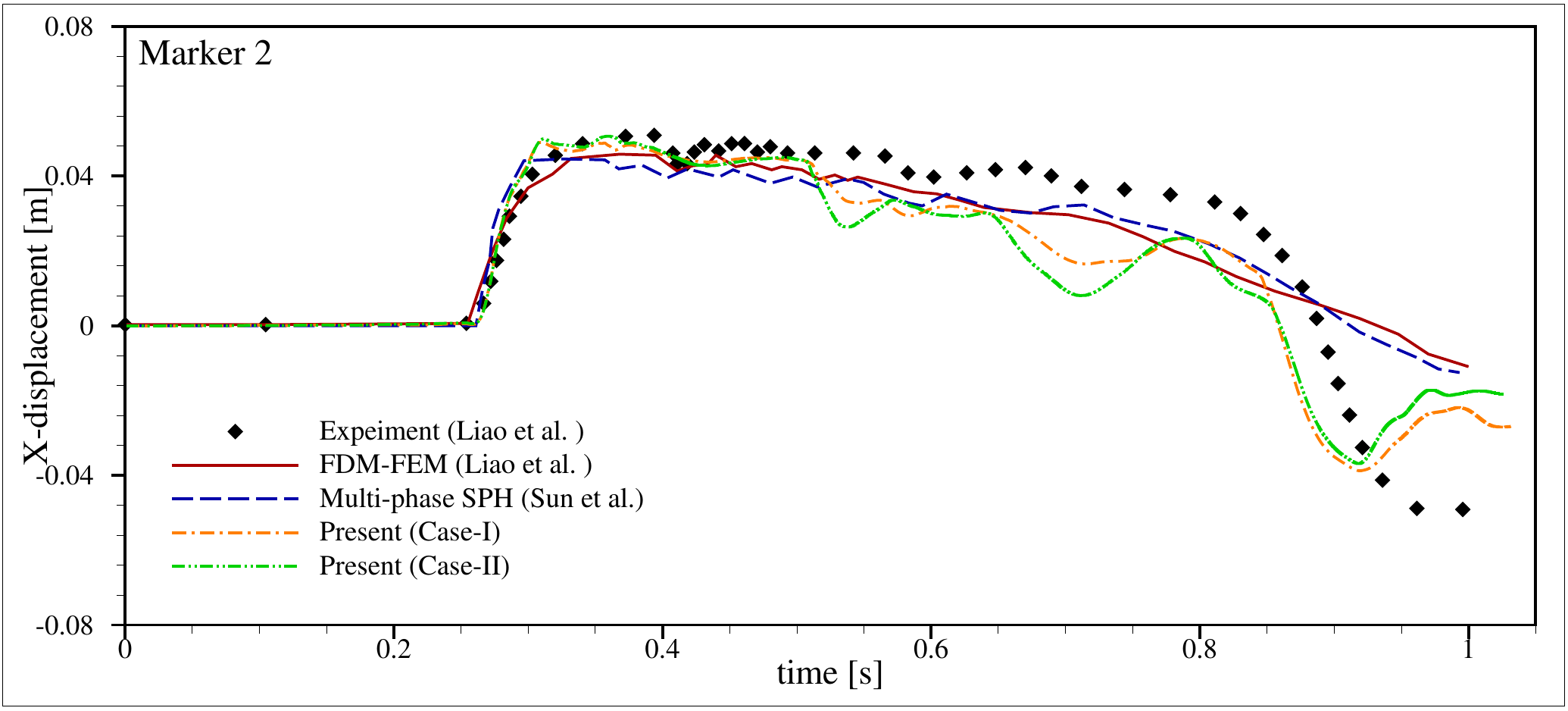}
	\includegraphics[trim = 1mm 1mm 1mm 1mm, clip, width=0.85\textwidth]{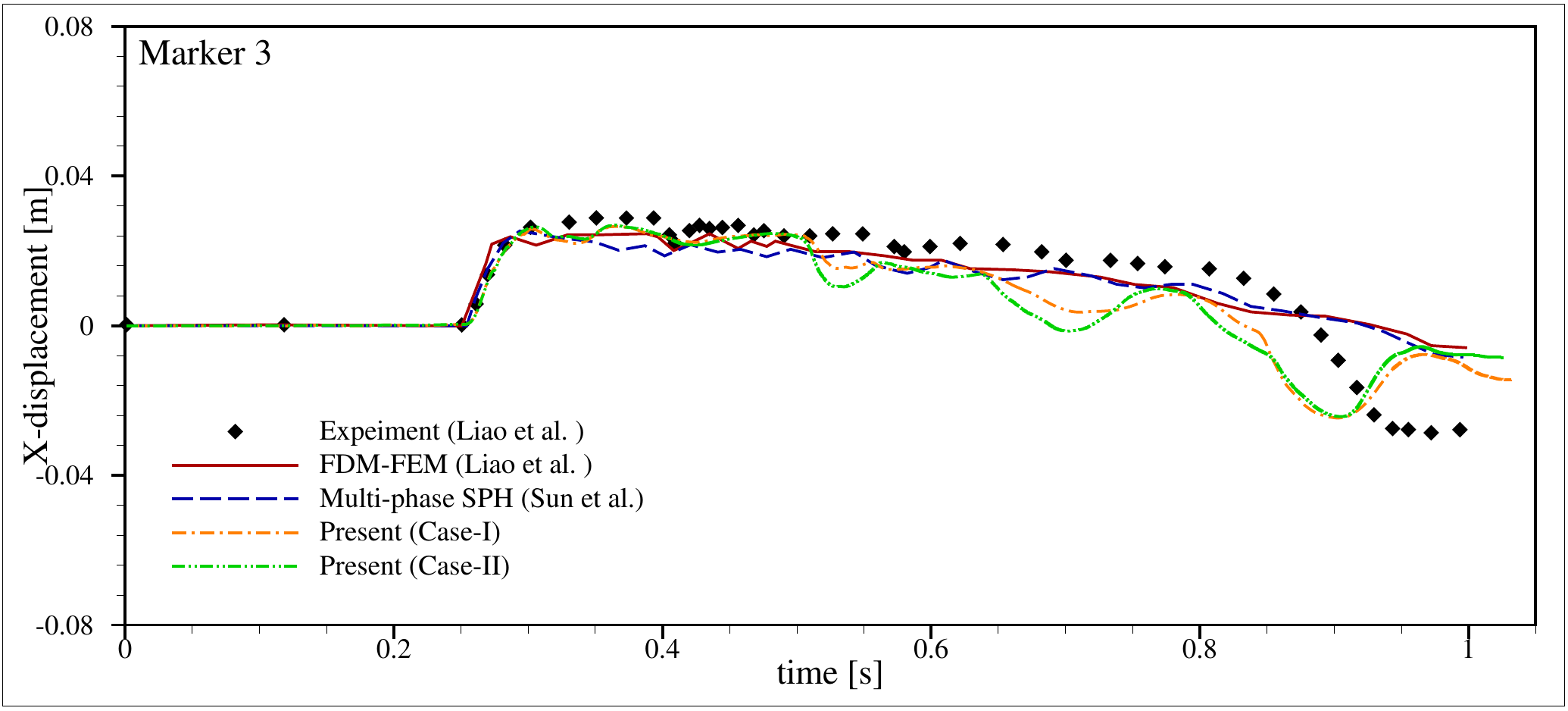}
	\caption{
		Dam-break flow impacting an elastic plate: 
		The time history of the horizontal displacement of the plate measured by Marker $1$, $2$ and $3$  
		and its comparison with experimental data and FDM-FEM prediction from Ref. \cite{liao2015free} 
		and multi-phase $\delta-$SPH result of Ref. \cite{sun2019study}. 
		(For interpretation of the references to color in this figure legend, the reader is referred to the web version of this article.)
	}
	\label{figs:mp-dam-fsi-data}
\end{figure}
To quantitatively validate accuracy, 
Figure \ref{figs:mp-dam-fsi-data} reports the time history of the horizontal displacement of the plate probed by  
Marker $1$, $2$ and $3$ for Case-I and Case-II, 
and the comparison against experimental data \cite{liao2015free} 
and numerical prediction obtained with FDM-FEM method \cite{liao2015free} and multi-phase $\delta-$SPH method \cite{sun2019study}.  
In general, 
the present method in both single- and multi-resolution scenarios 
demonstrates a good agreement with both the experimental and numerical data \cite{liao2015free, sun2019study} at the first impacting stage. 
At this stage, 
the displacement rapidly increases to its maximum value under the impulsive impacting force. 
As pointed out by Liao \etal \cite{liao2015free}, 
this stage is dominated by the first mode of vibration. 
However, 
discrepancy is observed afterward, 
for example 
the displacement between time $t = 0.35~\mathrm{s}$ and $t = 0.85~\mathrm{s}$ at Marker $1$ is underestimated 
by both the present and previous \cite{sun2019study} multi-phase SPH methods compared with the experimental observation \cite{liao2015free}. 
Compared with the those reported by Refs. \cite{sun2019study, liao2015free} with FDM-FEM or multi-phase $\delta-$SPH method, 
the present one can reasonably capture the higher modes of vibration (as shown in the zoom-in view of Figure \ref{figs:mp-dam-fsi-data}) and  
the backward displacement of the elastic plate observed at the final stage of the experiment, 
while slight difference is noted. 
It can be concluded that the present method and those in literature \cite{sun2019study, liao2015free} 
can well capture the first mode of vibration at the first impact stage, 
while the present one reasonably reproduce the the higher modes of vibration 
and the backward deflection with slight discrepancies likely to be associated with the stochastic nature of the impact pressures, 
the lack of exact repeatability of the experiment and the effect of the roughness of the boundary.  
It is worth noting that the present single- and multi-resolution simulations predict almost identical 
deflection of the plate, 
except slight difference in the higher modes vibration which exhibits large amplitude in the single resolution results, 
implying the air phase has a strong influence on it.

Having the above qualitative and quantitative validations, 
we can conclude that the present SPH method has demonstrated 
its robustness and accuracy in multi-resolution scenario for capturing large structure deformation under strong impact, 
meanwhile achieves improved computational speedup of $5.8$ compared with single-resolution counterpart as shown in Table \ref{tab:damfsi-cputime}.
\begin{table}[htb!]
	\centering
	\caption{Dam-break flow impacting an elastic plate: Analysis of computational efficiency. 
		The computer information is reported in Table \ref{tab:hydrostatic-fsi-cputime}. 
		For this test,  
		we evaluate the CPU wall-clock time for computation until $1.0~\mathrm s$ time instant.}
	\begin{tabular}{ccccc}
		\hline
		Cases     & 	\begin{tabular}{@{}c@{}}
			Multi resolution \\  
			(Case-I)
		\end{tabular} & \begin{tabular}{@{}c@{}}
		Single resolution \\  
		(Case-II)
	\end{tabular} & Speedup \\
		\hline
		CPU time (s)	&   421.02 & 2460.15 & 5.84     \\ 
		\hline
	\end{tabular}
	\label{tab:damfsi-cputime}
\end{table}
%
%
%
\section{Concluding remarks}\label{sec:conclusion}
In this paper, 
the multi-resolution SPH framework proposed by 
Zhang \etal~\cite{zhang2021multi} is extended to multi-phase flow and hydroelastic FSI.  
To cooperate with this multi-resolution framework, 
a simple and efficient multi-phase SPH method is first proposed  
by introducing different density reinitialization strategies other than 
realizing mass conservation through different formulations as Ref. \cite{rezavand2020weakly}. 
Then the transport velocity formulation are rewritten 
by applying localized background pressure to decrease the numerical dissipation. 
Also,  
the solid boundary condition proposed by Zhang \etal \cite{zhang2022} is adopted in the multi-resolution scenario 
to capture the FSI coupling. 
With several benchmark tests with high density ratio and complex phase interface, 
e.g., 
hydrostatic test, sloshing flow and dam-break flow, and its interaction with rigid and flexible structures, 
the present method demonstrates its the efficiency, accuracy and robustness. 
The performance of the present multi-resolution framework 
renders it a potential and powerful alternative in terms of computational efficiency for multi-physics applications 
in studying natural phenomena and engineering problems with proper surface tension and wetting model 
which is the main objective of our future work. 
%
%
\section*{CRediT authorship contribution statement}
{\bfseries Chi Zhang:} Conceptualization, Methodology, Investigation, Visualization, Validation, Formal analysis, Writing - original draft, Writing - review \& editing; 
{\bfseries Yujie Zhu:} Investigation, Formal analysis, Writing - review \& editing; 
{\bfseries Xiangyu Hu:} Investigation, Supervision. 
%
%
\section*{Declaration of competing interest }
The authors declare that they have no known competing financial interests 
or personal relationships that could have appeared to influence the work reported in this paper.
%
%
\section*{Acknowledgement}
The authors would like to express their gratitude to Deutsche Forschungsgemeinschaft (DFG) 
for their sponsorship of this research under grant number DFG HU1527/12-4. 
%
%
\bibliography{mybibfile}
%
%
\end{document}